\documentclass[12pt]{article}
\usepackage{latexsym,amssymb}
\newcommand{\be}{\begin{equation}}
\newcommand{\ee}{\end{equation}}
\newcommand{\beqs}{\begin{eqnarray}}
\newcommand{\eeqs}{\end{eqnarray}}
\def\vv{{\cal V}}
\def\({\left(}
\def\){\right)}

\newcommand{\Exc}[1]{{${\rm E}_{{#1}({#1})}$}}

\def\mxth{\mathsurround=0pt }
\def\xversim#1#2{\lower2.pt\vbox{\baselineskip0pt \lineskip-.5pt
x  \ialign{$\mxth#1\hfil##\hfil$\crcr#2\crcr\sim\crcr}}}

\def\lagr{{\cal L}}

\renewcommand{\a}{\alpha}
\renewcommand{\b}{\beta}

\renewcommand{\d}{\delta}
\newcommand{\pa}{\partial}
\newcommand{\g}{\gamma}

\newcommand{\e}{\epsilon}

\newcommand{\m}{\mu}
\newcommand{\n}{\nu}

\newcommand{\nn}{\nonumber}

\def\be{\begin{equation}}
\def\ee{\end{equation}}
\def\bea{\begin{eqnarray}}
\def\eea{\end{eqnarray}}

\newcommand{\ft}[2]{{\textstyle\frac{#1}{#2}}}

\newcommand{\eqn}[1]{(\ref{#1})}
\def\bfone{\relax{\rm 1\kern-.35em 1}}

\begin{document}
\begin{titlepage}
\begin{flushright}
ITP-UU-02/58 \\
SPIN-02/36\\[1mm]
{\tt hep-th/0212239}
\end{flushright}
\vskip 20mm
\begin{center}
{\Large {\bf On Lagrangians and Gaugings
}}\\[2mm]
{\Large {\bf of Maximal Supergravities }}

\vskip 10mm

{\bf Bernard de Wit, Henning Samtleben and Mario Trigiante}

\vskip 4mm

{\em Institute for Theoretical Physics \& Spinoza Institute}\\
{\em Utrecht University, Utrecht, The Netherlands}\\[2mm]

{\tt B.deWit@phys.uu.nl} \\
{\tt H.Samtleben@phys.uu.nl}\\
{\tt M.Trigiante@phys.uu.nl}

\vskip 6mm

\end{center}

\vskip .2in

\begin{center} {\bf Abstract } \end{center}
\begin{quotation}\noindent
A consistent  gauging of maximal supergravity requires that the
$T$-tensor transforms according to a specific representation of the
duality group. The analysis of viable gaugings is thus amenable to
group-theoretical analysis, which we explain and exploit for a
large variety of gaugings. We discuss the subtleties in four spacetime
dimensions, where the ungauged Lagrangians are not unique and encoded
in an ${\rm E}_{7(7)}\backslash {\rm Sp}(56;\mathbb{R})/{\rm GL}(28)$ 
matrix. Here we define the $T$-tensor and derive all relevant
identities in full generality. We present a large number of examples in
$d=4,5$ spacetime dimensions which include non-semisimple gaugings
of the type arising in (multiple) Scherk-Schwarz reductions. We also
present some general background material on the latter as well
as some group-theoretical results which are necessary for using
computer algebra. 
\end{quotation}

\end{titlepage}

\eject

\section{Introduction}
Maximal supergravity theories contain a number of vector gauge fields
which have an optional coupling to themselves as well as to other
supergravity fields. The corresponding gauge groups are nonabelian. To
preserve supersymmetry in the presence of these gauge couplings, the
Lagrangian must contain masslike terms for the fermions and a
potential depending on the scalar fields. In nonmaximal supergravity
these terms are often described by means of auxiliary fields and/or
moment maps; in the maximally supersymmetric theories the effect of
the masslike terms and the potential is encoded in the so-called
$T$-tensor \cite{deWitNic}. It is a subtle matter to determine which
gauge groups and corresponding charge assignments are compatible with
supersymmetry. Based on Kaluza-Klein compactifications of
higher-dimensional maximal supergravities on spheres, one readily
concludes that the gauge groups ${\rm SO}(8)$, ${\rm SO}(6)$ and ${\rm
SO}(5)$ are possible options in $d=4,5$ and 7 dimensions,
respectively, corresponding to the isometry groups of $S^7$, $S^5$ and
$S^4$ \cite{deWitNic,GunaRomansWarner,PPvN}. But also noncompact and
non-semisimple groups turn out to be possible
\cite{Hull,GunaRomansWarner,AndCordFreGual}, which are noncompact
versions and/or contractions of the orthogonal groups. More recent
work revealed the so-called `flat' gauge groups that one obtains upon
Scherk-Schwarz reductions of higher-dimensional theories
\cite{ScherkSchwarz}, as well as several other non-semisimple groups
\cite{AndDauFerrLle,Hull2,exhaustive}.  In $d=3$ dimensions there is
no guidance from Kaluza-Klein compactifications and one must rely on
a group-theoretical analysis \cite{NicSam2}. In this paper we apply
the same kind of analysis to gaugings in higher dimensions.

Apart from the choice of the gauge group, a number of other subtleties
arise that depend on the number of spacetime dimensions. In $d=3$
dimensions supergravity does not contain any vector fields, because
these can be dualized to scalar fields. Nevertheless a gauging can be
performed by introducing vector fields via a Chern-Simons term (so
that new dynamic degrees of freedom are avoided), which are
subsequently coupled to some of the ${\rm E}_{8(8)}$ invariances of
the supergravity Lagrangian~\cite{NicSam}. In that case there exists a
large variety of gauge groups of rather high dimension. In $d=4$
dimensions there are 28 vector gauge fields, but the ${\rm E}_{7(7)}$
invariance is not reflected in the Lagrangian but only in the combined
field equations and Bianchi identities by means of electric-magnetic
duality. This duality rotates magnetic and electric charges, but the
gauge couplings must be of the electric type. Then, in $d=5$
dimensions, tensor and vector gauge fields are dual to one another in
the absence of charges. The ${\rm E}_{6(6)}$ invariance is only
manifest when all the tensor fields have been converted to vector
fields (transforming according to the 27-dimensional
representation). In the presence of charges, however, the vector
fields must either correspond to a nonabelian gauge group or they must
be neutral. Charged would-be vector fields that do not correspond to
the nonabelian gauge group, should be converted into antisymmetric
tensor fields~\cite{TownPilcNieuw}. This implies that the field
content of the $d=5$ theory depends on the gauge group.

The Lagrangian of ungauged maximal supergravity contains the standard
Einstein-Hilbert, Rarita-Schwinger and Dirac Lagrangians for the
gravitons, the gravitini and the spinor fields. The kinetic terms of
the gauge fields depend on the scalar fields and the kinetic term for
the scalar fields takes the form of a nonlinear sigma model based on a
symmetric coset space G/H. Here H is the maximal compact subgroup of
G; a list of these groups is given in table~\ref{maximal-cosets}. The
Lagrangian (or the combined field equations and Bianchi identities) is
invariant under the isometry group G which is referred to as the
duality group. The standard treatment of gauged nonlinear sigma models
exploits a formulation in which the group H is realized as a local
invariance which acts on the spinor fields and the scalars; the
corresponding connections are composite fields. The gauging is based
on a gauge group ${\rm G}_g\subset {\rm G}$ whose connections are
(some of the) elementary vector gauge fields of the supergravity
theory. The matrix which encodes the embedding of the gauge group into
the duality group is in fact linearly related to the $T$-tensor. The
coupling constant associated with the gauge group ${\rm G}_g$ will be
denoted by $g$. One can impose a gauge condition with respect to the
local H invariance which amounts to fixing a coset representative for
the coset space. In that case the G-symmetries will act nonlinearly on
the fields and these nonlinearities make many calculations intractable
or, at best, very cumbersome. Because it is much more convenient to
work with symmetries that are realized linearly, the best strategy is
therefore to postpone the gauge fixing till the end.

This paper aims at exploiting the group-theoretical constraints on the
$T$-tensor, which are essential in order to have a consistent,
supersymmetric gauging. It is well-known that the $T$-tensor must be
restricted to a certain representation of the duality group. For
instance, in four dimensions, this is the ${\bf 912}$ representation
of \Exc7, and in five dimensions it is the ${\bf 351}$ representation
of \Exc6. We derive these allowed representations for
dimensions $d=3,\dots, 7$. Possible gaugings can then be explored by
investigating which gauge groups lead to $T$-tensors that belong to
the required representation. This proves to be sufficiently powerful
to completely identify the possible gauge groups within a given
subgroup of G, and to determine which gauge fields and generators of G
are involved in each of the gaugings.  Part of the analysis is done
with help of the computer. To demonstrate the method and its
potential, we analyze a number of gaugings in $d=4, 5$ dimensions,
including the known cases. Applications with hitherto unknown gauge
groups are relegated to a forthcoming publication~\cite{dWSamtTrig2}.

In four dimensions the Lagrangian is not unique in the absence of
charges, because of electric/magnetic duality. Once the charges are
switched on, the possibility of obtaining alternative Lagrangians is
restricted, because electric charges cannot be converted to magnetic
ones. Without introducing the gauging, there exist different
Lagrangians ({\it i.e.}, not related by local field redefinitions)
with different symmetry groups, whose field equations and Bianchi
identities are equivalent and share the same invariance group. This
feature makes the four-dimensional case more subtle to analyze and
therefore considerable attention is given to this case. In particular,
we show that the different Lagrangians of the ungauged theory are
encoded in a matrix ${\sf E}$ belonging to ${\rm E}_{7(7)}\backslash
{\rm Sp}(56;\mathbb{R})/{\rm GL}(28)$.

Some of the gaugings can be interpreted as originating from a
Scherk-Schwarz truncation of a higher-dimensional theory
\cite{ScherkSchwarz}. In order to identify such gaugings we have
included some material on these reductions and we exhibit examples in
four and five dimensions. Both of them are single reductions,
originating from a theory with one extra dimension. However, also
multiple reductions are possible from theories with more than one
extra dimensions, which lead to more complicated gauge groups, as we
shall discuss in more detail in~\cite{dWSamtTrig2}.

This paper is organized as follows. In section~2 we review the
structure of the nonlinear sigma models that appear in maximal
supergravity and the symmetries of the Lagrangians. In section~3 we
focus on the definition of the $T$-tensor in four dimensions and
discuss a large number of relevant features. In section~4, we derive
the group theoretical constraint on the $T$-tensor (and equivalently
on the embedding matrix of the gauge group) for dimensions $d=3,
\dots, 7$. This constraint provides an efficient criterion for
identifying consistent gaugings. In section~5 we review characteristic
features of Scherk-Schwarz reductions, which correspond to some of the
gaugings. Finally, in sections~6 and 7, we demonstrate how this
framework naturally comprises the known gaugings in $d=4$ and 5
dimensions. In addition, we exhibit a new gauging in $d=5$ dimensions
which can be interpreted as a Scherk-Schwarz reduction. An appendix is
included with some group-theoretical results.

\section{Coset-space geometry and duality}
\setcounter{equation}{0}
\label{coset}
In this section we review the coset-space structure of the maximally
supersymmetric supergravity theories. Because of the subtleties of
supergravity in four spacetime dimensions special attention is devoted
to this theory. In particular, we discuss its inequivalent
Lagrangians, encoded in a matrix ${\sf E}$. The existence of different
Lagrangians makes the analysis of the various gaugings more
complicated, as they are associated with different classes of
gaugings.

The scalar fields in maximal supergravity parametrize a
symmetric G/H coset space and the theory is realized with a local H
symmetry with composite connection fields. The standard formalism
starts from a matrix-valued field, $\vv(x)$, that belongs to the group
G, usually in the fundamental representation. Before introducing an
(optional) gauging, this field
transforms under rigid G transformations from the left and
under local H transformations from the right. G-invariant one-forms
are defined by
\be
\label{Q-P}
\vv^{-1}\partial_\mu\vv = {\cal Q}_\m + {\cal P}_\m \,,
\ee
where ${\cal Q}_\m$ and ${\cal P}_\m$ take their values in the Lie
algebra associated with G; ${\cal Q}_\mu$ acts as a gauge field
associated with the local
$\rm H$ transformations. Eventually one may fix the gauge
freedom associated with H, but until that point $\vv$ will just be an
unrestricted spacetime dependent element of the group G. After
imposing the gauge condition on $\vv(x)$ one obtains the coset
representative $\vv(\phi(x))$, where the fields $\phi(x)$ parametrize
the coset space. The spinor fields of the
supergravity Lagrangian transform under H, but are invariant under
G. It is convenient to work with an H-covariant derivatives, which on
$\vv$ is equal to $D_\mu \vv = \partial_\mu \vv - \vv {\cal Q}_\m$,
so that \eqn{Q-P}  can be written as,
\be
\vv^{-1}D_\m\vv ={\cal P}_\m  \,.  \label{Dvv}
\ee

The quantities ${\cal Q}_\m$ and ${\cal P}_\m$ are subject to the
Cartan-Maurer equations, which follow directly from
\eqn{Q-P},
\bea
F_{\mu\nu}({\cal Q})& = & \partial_\mu {\cal Q}_\nu - \partial_\nu
{\cal Q}_\mu +[ {\cal Q}_{\mu}, {\cal Q}_{\nu }]  =  - [{\cal
P}_{\mu},  {\cal P}_{\nu }]  \,,  \nonumber \\
D_{[ \mu} {\cal P}_{\nu]} & = & \pa_{[\m}{\cal P}_{\n]} + [
{\cal Q}_{[\m}, {\cal P}_{\n]}] = 0 \,. \label{CM-Q-P}
\eea
Here we made use of the fact that the generators of the
subgroup H close and that the remaining generators
associated with G/H form a representation of the group H. Furthermore,
the commutators of any two of the latter generators are proportional
to the generators of H. This last requirement is responsible for the
zero on the right-hand side of the second
Cartan-Maurer equation and ensures that we are dealing with a
symmetric coset space.
The Lagrangian of the corresponding nonlinear sigma model is
invariant under both rigid G transformations and local H
transformations and reads,
\be
\lagr \propto \ft12 {\rm tr} \, \Big[D_\mu \vv^{-1}\, D^\mu \vv \Big]
=   - \ft{1}{2} {\rm tr}\,\Big[ {\cal P}_\mu \,{\cal P}^\mu\Big] \,.
\label{eq:AA-lagrangian}
\ee

\begin{table}[tb]
\begin{center}
\begin{tabular}{l l l l}\hline
$d$ &G        & H   & ${\rm dim}\,[{\rm G}]-{\rm dim}\,[{\rm H}]$ \\
\hline
11  & 1       & 1   & $0-0=0$      \\
10A & SO$(1,1)/{\bf Z}_2$   & 1   & $1-0=1$  \\
10B & SL(2)    & SO(2) &  $3-1=2$  \\
9   & GL(2)    & SO(2) &  $4-1=3$ \\
8   & E$_{3(+3)}\sim {\rm SL}(3)\! \times \!{\rm SL}(2)$    &
U(2) & $11- 4=7$  \\
7   & E$_{4(+4)}\sim {\rm SL}(5)$  & USp(4) &$24-10=14$   \\
6   & E$_{5(+5)}\sim {\rm SO}(5,5)$ &
    USp$(4)\!\times\! {\rm USp}(4)$ &$45-20=25$   \\
5   & E$_{6(+6)}$  & USp(8) & $78-36= 42$    \\
4   & E$_{7(+7)}$  &  SU(8) & $133-63= 70$    \\
3   & E$_{8(+8)}$  & SO(16) & $248 - 120= 128$   \\ \hline
\end{tabular}
\end{center}
\caption{\small
Homogeneous scalar manifolds G/H for maximal
supergravities in various dimensions. The type-IIB theory
cannot be obtained from reduction of 11-dimensional supergravity
and is included for completeness. The difference of the
dimensions of G and H equals the number of scalar fields.
}\label{maximal-cosets}
\end{table}

For maximal supergravity theories the symmetric cosets are known
\cite{cremmer3}. For the convenience of the reader we have listed them
in table~\ref{maximal-cosets}.  Part of the isometry group can now be
gauged by coupling the (elementary) vector gauge fields $A_\m$ to a
subset of the generators corresponding to a group ${\rm G}_g\subset
{\rm G}$. The dimension of this gauge group is restricted by the
number of available vector gauge fields.  We have listed the field
content for the bosonic fields assigned to representations of H in a
second table~\ref{maximal-sg-bosons}. Because the gauge group ${\rm
G}_g$ is embedded in the isometry group G, it must act on $\vv$ so
that the covariant derivative of $\vv$ changes by the addition of the
gauge fields $A_\m$ which take their values in the Lie algebra
corresponding to ${\rm G}_g$,
\begin{equation}
  \label{eq:gauged-cov-der}
D_\mu \vv(x) = \partial_\mu \vv(x) - \vv(x)\, {\cal Q}_\m(x)  -
g\,A_\m(x)  \,\vv(x) \,.
\end{equation}
With this change, the expressions for ${\cal Q}_\m$ and ${\cal P}_\m$
are still given by \eqn{Dvv}, but the derivative is now covariantized
and modified by the terms depending on the new gauge fields
$A_\m$. The consistency of this procedure is obvious as \eqn{Dvv} is
fully covariant. Of course, the original rigid invariance under G
transformations from the left is in general broken by the embedding of
the new gauge group ${\rm G}_g$ into G.

The modifications caused by the new minimal couplings are minor and
the effects can be concisely summarized by the Cartan-Maurer
equations,
\bea
 \label{CM-gauged}
{\cal F}_{\mu\nu}({\cal Q}) &=&
[{\cal  P}_{\mu}, {\cal P}_{\nu}]  -g \Big[\vv^{-1}
F_{\m\n}(A)\vv\Big]_{\rm H} \,,
\nonumber \\
D_{[ \mu} {\cal P}_{\nu]} & = & -\ft12 g \Big[\vv^{-1}
F_{\m\n}(A)\vv\Big]_{\rm G/H} \,.
\eea
Note that ${\cal P}_\m$ and ${\cal Q}_\m$ are invariant under ${\rm
G}_g$ (but transform under local H-transformat\-ions, as before).

\begin{table}[htb]
\begin{center}
\begin{tabular}{l l l l l l l  }\hline
$d$ &${\rm H}_{\rm R}$& $p=0$ & $p=1$ & $p=2$ & $p=3$ &
$p=4\!$  \\   \hline
11  & 1      & 0   & 0   & 0  & 1  & 0 \\[.5mm]
10A$\!\!\!$ & 1  & 1   & 1   & 1  & 1  & 0 \\[.5mm]
10B$\!\!\!$ & SO(2)  & 2   & 0   & 2  & 0  & $1^\ast$ \\[.5mm]
9   & SO(2)  &$2+1$&$2+1$& 2  & 1  & ~ \\[.5mm]
8   & U(2)   & $5+1+\bar 1\!$ & $3+\bar 3$ &3 &$[1]$&~\\[.5mm]
7   & USp(4) & 14  & 10  & 5  & ~  & ~ \\[.5mm]
6   &USp$(4)\times{\rm USp}(4)$ &(5,5)&(4,4)&$(5,1)+(1,5)$ & ~ & ~\\
5   & USp(8) & 42 & 27   & ~  & ~  & ~ \\[.5mm]
4   & U(8)   & $35+\overline{35}$ & $[28]$ & ~ & ~ & ~ \\[.5mm]
3   & SO(16) & 128 & ~   & ~  & ~  & ~ \\ \hline
\end{tabular}
\end{center}
\caption{\small
Bosonic field content for maximal supergravities
described by $p$-rank antisymmetric gauge fields; $p=0$ corresponds to
a scalar field and the graviton fields has been suppressed. The $p=4$
gauge field in $d=10$B has a self-dual field strength. The
representations [1] and [28] (in $d=8,4$, respectively) are extended
to U(1) and SU(8) representations through duality transformations on
the field strengths. These transformations can not be represented on
the vector potentials. In $d=3$ dimensions, the graviton does not
describe propagating degrees of freedom. For $p>0$ the fields can be
assigned to representations of a bigger group than ${\rm H}_{\rm R}$.
}\label{maximal-sg-bosons}
\end{table}

In the following we concentrate on $d=4$ spacetime dimensions
\cite{cremmer,deWitNic}, where the Lagrangian is not uniquely defined
and alternative Lagrangians, not related by local field redefinitions,
can be obtained (in the absence of charges) via so-called
electric-magnetic duality transformations (for a recent review, see
\cite{susy30}). These transformations constitute the group ${\rm
Sp}(56;\mathbb{R})$. Lagrangians related via electric-magnetic duality
do not share the same symmetry group and therefore they may allow
different gaugings as the gauge group must be embedded into this
group. Once the charges have been switched on, the possibilities for
performing electric-magnetic duality are severely restricted, as
electric charges cannot be converted to magnetic ones via local field
redefinitions. This complication is specific to 4 dimensions; for
$d\not=4$ the situation is simpler and the results of this section can
be taken over without much difficulty.

For $d=4$, $\vv(x)$ is a $56\times 56$ matrix, sometimes called the
56-bein, which decomposes as follows,
\begin{equation}
  \label{eq:u-v}
\vv(x) = \pmatrix{ u^{ij}{}_{\!IJ}(x) & -v_{ kl  IJ}(x)\cr
\noalign{\vskip 6mm}
-v^{ijKL}(x) &   u_{kl}{}^{\!KL}(x)\cr} \,.
\end{equation}
The indices $I,J,\ldots$ and $i,j,\ldots$ take the
values $1,\ldots,8$, so that there are 28 antisymmetrized index pairs
representing the matrix indices of $\vv$; the row indices are
$([IJ],[KL])$, and the column indices are $([ij],[kl])$, so as to
remain consistent with the conventions of \cite{deWitNic}. The
above matrix is pseudoreal and belongs to ${\rm
E}_{7(7)}\subset {\rm Sp}(56;\mathbb{R})$ in the fundamental
representation.\footnote{
  The pseudoreal representation stresses the maximal compact ${\rm
  SU}(8)$ subgroup. It would perhaps be appropriate to denote the
  pseudoreal representation by ${\rm USp}(28,28)$, but for reasons of
  uniformity we will always refer to ${\rm Sp}(56;\mathbb{R})$. }
We use the convention where
$u^{ij}{}_{\!IJ}= (u_{ij}{}^{\!IJ})^\ast$ and $v_{ijIJ}=(v^{ij
IJ})^\ast$.  The indices $i,j,\ldots$ refer to ${\rm SU}(8)$ and
capital indices $I,J,\ldots$ are subject to \Exc7 transformations.
Using the above definition, one may evaluate the quantities ${\cal
Q}_\m$ and ${\cal P}_\m$,
\be
\label{P-Q-max}
\vv^{-1} \pa_\m\vv = \pmatrix{{\cal Q}_{\m\, ij}{}^{\!mn}
&{\cal P}_{\m\, ijpq} \cr \noalign{\vglue 6mm}
{\cal P}_\m^{klmn} & {\cal Q}_\m{}^{\!kl}{}_{\!pq} \cr}\,,
\ee
which leads to the expressions,
\bea
{\cal Q}_{\m \,ij}{}^{\!kl}&=& u_{ij}{}^{\!IJ}\, \pa_\m u^{kl}{}_{\!IJ}
-v_{ijIJ}\, \pa_\m v^{klIJ} \,, \nonumber\\
{\cal P}_\m^{ijkl} &=&  v^{ijIJ}\,\pa_\m u^{kl}{}_{\!IJ}-
u^{ij}{}_{\!IJ} \, \pa_\m v^{klIJ}  \,.
\eea
Compatibility with the Lie algebra of {\rm \Exc7} implies that ${\cal
P}_\m^{ijkl}$ is a selfdual ${\rm SU}(8)$ tensor,
\be
{\cal P}_\m^{ijkl} = \ft1{24}\,\varepsilon^{ijklmnpq}\, {\cal
 P}_{\m\,mnpq}\,,
\ee
and ${\cal Q}_\m$ transforms as a connection associated with
${\rm SU}(8)$. Hence, ${\cal Q}_{\m \,ij}{}^{\!kl}$ satisfies
the decomposition,
\be
{\cal Q}_{\m\, ij}{}^{\!kl}= \d^{[k}_{[i}\, {\cal Q}_{\m\,
j]}{}^{\!l]}\,,
\ee
with
\be
{\cal Q}_{\m \,i}{}^{\!j} = \ft23 \Big[u_{ik}{}^{\!IJ}\, \pa_\m
u^{jk}{}_{\!IJ} - v_{ikIJ}\, \pa_\m v^{jkIJ}  \Big]\,,
\ee
and ${\cal Q}_{\m}{}^{\!i}{}_{\!j} = - {\cal
Q}_{\m j}{}^i$ and ${\cal Q}_{\m i}{}^i=0$.

While the index pairs $[IJ]$ refer to the row indices of $\vv$ and are
subject to {\rm \Exc7}, the 28 gauge fields $A_\m^{AB}$ are labelled
by index pairs $[AB]$, where $A,B= 1,\ldots, 8$. As it turns out
\cite{deWit02}, the ungauged Lagrangians can be encoded into a matrix
${\sf E}$ belonging to ${\rm E}_{7(7)}\backslash {\rm
Sp}(56;\mathbb{R})/{\rm GL}(28)$, which defines the embedding of the
28 vector fields into the 56-bein and thus connects the two types of
index pairs $[IJ]$ and
$[AB]$,\footnote{
Similar additional parameters in four-dimensional Lagrangians have
been exploited also in $N=2,4$ supergravity \cite{deRooW,Trigetal}. }
\be
{\sf E}=\pmatrix{{\sf U}_{IJ}{}^{\!AB} & {\sf V}_{IJCD}\cr
\noalign{\vskip4mm}  {\sf V}^{KLAB}& {\sf U}^{KL}{}_{\!CD} \cr}\,.
\ee
Two Lagrangians related by electric-magnetic duality correspond to two
matrices ${\sf E}$ related by multiplication from the left by an
element of ${\rm Sp}(56;\mathbb{R})$. These matrices are not unique,
because an ${\rm E}_{7(7)}$ transformation can always be absorbed into
the 56-bein and a ${\rm GL}(28;\mathbb{R})$ transformation can be
absorbed into the gauge fields. It is convenient to include $\sf E$
into the 56-bein according to,
\be
\label{u-v-mod}
\hat\vv(x) = {\sf E}^{-1}\,\vv(x) \;,
\ee
where we have to remember that $\hat \vv$ is now no longer a group
element of ${\rm E}_{7(7)}$! This definition leads to corresponding
submatrices $u^{ij}{}_{\!AB}$ and $v^{ijAB}$.

Although the ${\rm E}_{7(7)}$ tensors ${\cal Q}_\m$ and ${\cal P}_\m$
are not affected by the matrix $\sf E$ and have identical expressions in
terms of $\vv$ and $\hat \vv$, the remaining interactions depend on
$\sf E$, and so do the transformation rules. However, by making use of
\eqn{u-v-mod} we can make the dependence on ${\sf E}$ implicit,
provided we also introduce an ${\rm SU}(8)$ (selfdual) covariant field
strength, defined by
\be
\label{su8-fs}
 F_{\m\n}^{+AB}= (u^{ij}{}_{\!AB}+ v^{ijAB}) \,{\sf F}_{\m\n ij}^{+}
-  (u_{ij}{}^{\!AB}+ v_{ijAB}) \, {\cal O}_{\m\n}^{+ij}\,,
\ee
where $F^{AB}_{\m\n} = 2\pa_{[\m} A_{\n]}^{AB}$, and ${\cal
O}_{\m\n}^{+ij}$ is a selfdual Lorentz tensor that comprises terms
quadratic in the fermion fields. The anti-selfdual tensors are
obtained by complex conjugation. The terms in the Lagrangian that
depend on the field strengths, take the form
\bea
\label{L3}
\lagr\!&=& \!-\ft18 e\, {\cal N}_{AB,CD}\, F^{+AB}_{\m\n} \,
F^{+CD\,\m\n}  -\ft12 e\, F^{+AB}_{\m\n} \,
[(u+v)^{-1}]^{AB}{}_{\!\!ij}\, 
{\cal O}^{+\m\n\,ij}    \nn\\[1mm]
&& + \mbox{ h.c.}\,,
\eea
where the complex $28\times 28$ symmetric matrix ${\cal N}$ is defined
by $(u^{ij}{}_{\!AB} +v^{ijAB}) \,{\cal N}_{AB,CD} = u^{ij}{}_{\!CD}
-v^{ijCD}$. Obviously, \eqn{L3} depends only implicitly on ${\sf E}$.

The choice for ${\sf E}$ has a bearing on the manifest subgroup of
${\rm E}_{7(7)}$ under which the Lagrangian is invariant. We shall
call this group the {\it electric} duality group ${\rm G_e}$ and
define it as the largest subgroup of ${\rm E}_{7(7)}$ which acts on
all the fields, including the 28 gauge fields $A_\m^{AB}$, that leaves
the action invariant. Defining $G_{\mu\nu \,AB}\propto
\varepsilon_{\m\n\rho\sigma}\,
\pa{\lagr}/ \pa F_{\rho\sigma}^{AB}$ the group ${\rm G}_{\rm e}$ acts
as follows,
\bea
\label{spelec}
\d F_{\m\n}^{AB} &=& \tilde \Lambda^{AB}{}_{\!CD}
\,F_{\m\n}^{CD}\,,\nonumber \\
\d G_{\m\n\,AB}  &=& -\tilde \Lambda^{CD}{}_{\!AB} \,G_{\m\n\, CD}
+ \tilde\Sigma_{ABCD} \,F_{\m\n}^{CD}\,,
\eea
with $\tilde \Lambda^{AB}{}_{CD}$ and $\tilde\Sigma_{ABCD}=
\tilde\Sigma_{CDAB}$ real. Obviously, the $\tilde\Lambda$ characterize
the transformations of the vector potentials $A_\m^{AB}$; the part of
the generators that resides in $\tilde \Sigma$ is realized in the
transformations of $\vv$. The variations \eqn{spelec} generate the
subgroup ${\rm G_e}\subset {\rm E}_{7(7)}$.  However, we use a
formulation based on the complex combinations $(iG \pm F)_{\m\n}$
which is connected to the \Exc7 basis for $\vv$ via the matrix ${\sf
E}$. This is the basis that is relevant for $u^{ij}{}_{\!AB}$ and
$v_{ijAB}$, which transform according to
\begin{eqnarray}
\label{u-v-trans}
\delta u_{ij}{}^{\!AB}&=& - u_{ij}{}^{\!CD}\, \Lambda_{CD}{}^{\!AB}  -
v_{ijCD}\,\Sigma^{CDAB} \,,\nonumber\\
\delta v^{ijAB} &=&  -v^{ijCD}\,\Lambda_{CD}{}^{\!AB} -
 u^{ij}{}_{\!CD}\,\Sigma^{CDAB}\,,
\eea
where
\bea
\label{LambdaSigma}
\Lambda_{AB}{}^{\!CD} = (\Lambda^{AB}{}_{\!CD})^\ast &=& \ft12
(\tilde\Lambda^{AB}{}_{\!CD}-\tilde \Lambda^{CD}{}_{\!AB}  +i
\tilde \Sigma_{ABCD})\,,\nonumber \\
\Sigma _{ABCD} = (\Sigma^{ABCD})^\ast &=& -\ft12
(\tilde\Lambda^{AB}{}_{\!CD}+\tilde \Lambda^{CD}{}_{\!AB}  +i
\tilde \Sigma_{ABCD})\,.
\eea
Clearly these parameters are subject to the constraint
\be
\mbox{Im } \Big(\Sigma_{ABCD} + \Lambda_{AB}{}^{\!CD}\Big) = 0\,.
\label{uspelec}
\ee
Under these transformations we find the following results,
\bea
\label{sumelec} 
\delta (u_{ij}{}^{\!AB} + v_{ijAB}) &=& \tilde \Lambda^{AB}{}_{\!CD}\,
(u_{ij}{}^{\!CD} + v_{ijCD})\,
  \,,\nonumber\\ 
\delta (u^{ij}{}_{\!AB} - v^{ijAB}) &=& - \tilde \Lambda^{CD}{}_{\!AB}
\,(u^{ij}{}_{\!CD} - v^{ijCD})\, +\tilde\Sigma_{ABCD}\,
(u^{ij}{}_{\!CD} + v^{ijCD})   \,,\nonumber\\ 
\delta {\cal N}_{AB,CD} &=& -\tilde\Lambda^{EF}{}_{\!AB} \, 
{\cal N}_{EF,CD}  - {\cal N}_{AB,EF}\,\tilde\Lambda^{EF}{}_{\!CD} 
+ i\tilde\Sigma_{ABCD} \,. 
\eea
It is now easy to verify that the Lagrangian \eqn{L3} is not invariant
under these transformations but changes instead into a total
derivative,
\be
\label{total}
\d\lagr = \ft1{16}\varepsilon^{\m\n\rho\sigma}\, \tilde\Sigma_{ABCD} 
\;F^{AB}_{\m\n}\, F_{\rho\sigma}^{CD}\,.
\ee
Transformations with $\tilde \Sigma \neq 0$ induce a shift in the
generalized theta angle and are therefore called Peccei-Quinn
transformations. When the Peccei-Quinn transformations are part of a
nonabelian gauge group associated with the gauge fields $A_\m^{AB}$,
so that the corresponding $\tilde\Sigma$ depends on the spacetime
coordinates, then \eqn{total} is no longer a total derivative. In that
case one must include a Chern-Simons-like term
\be
\label{total-local}
\lagr \propto g\, \varepsilon^{\m\n\rho\sigma}\, \tilde\Sigma_{ABCD;EF}  
\;A_\m^{EF}\,A_\n^{AB}\, (\pa^{~}_\rho A_\sigma^{CD} -\ft18g\,
f_{GH,IJ}{}^{\!\!CD} A_\rho^{GH} A_\sigma^{IJ} )  \,,
\ee
where the constants $g \,\tilde\Sigma_{ABCD;EF}$ are the coefficients
that one obtains when expanding $\tilde\Sigma$ in terms of the gauge
group parameters $\Xi^{EF}(x)$, and the $f_{AB,CD}{}^{\!\!EF}$ are the
structure constants of the gauge group. The addition of the term
\eqn{total-local} not only restores the gauge invariance of the
action, but also the supersymmetry, as was shown in \cite{dWLVP}. The
fact that the constants $\tilde\Sigma_{ABCD;EF}$ emerge as variations
of the tensor ${\cal N}$ under the action of a nonabelian group,
implies that they are subject to certain constraints. These
constraints allow only nontrivial solutions when the group is
non-semisimple~\cite{dWHR}.

Without the condition \eqn{uspelec}, the variations
\eqn{u-v-trans} define an infinitesimal ${\rm
Sp}(56;\mathbb{R})$ transformation in the pseudoreal basis. In terms of
\eqn{spelec} the extra contributions are related to an extra
variation of $\delta F_{\m\n}\propto G_{\m\n}$ proportional to a
second real parameter $\tilde \Sigma^\prime$. Its effect is to 
shift the contribution of $\tilde \Sigma$ in the two equations in 
\eqn{LambdaSigma} by equal but opposite amounts, such that
\eqn{uspelec} is no longer satisfied. The maximal
compact subgroup of ${\rm Sp}(56;\mathbb{R})$ resides in the
$\Lambda_{AB}{}^{\!CD}$ and is equal to ${\rm U}(28)$.

In the following we distinguish some classes of Lagrangians with
inequivalent symmetry groups ${\rm G}_{\rm e}$ and corresponding
matrices ${\sf E}$. Different matrices ${\sf E}$ are related by
electric-magnetic duality transformations, which can be worked out
explicitly for any given case. We concentrate on the semisimple
subgroups of ${\rm G}_{\rm e}$. Obviously the group ${\rm G}_{\rm e}$
must have a {\it real} 28-dimensional representation. We will discuss
four inequivalent cases, based on the semisimple groups
${\rm SL}(8,\mathbb{R})$,
${\rm E}_{6(6)}\times {\rm SO}(1,1)$, ${\rm SL}(2,\mathbb{R})\times
{\rm SO}(1,1)\times {\rm SL}(6,\mathbb{R})$, and ${\rm SU^\star (8)}$.

\paragraph{The ${\rm SL}(8,\mathbb{R})$-basis: ${\sf E}={\sf E}_{{\rm
SL}(8,\mathbb{R})}={\rm Id}$}

The Lagrangian is invariant under ${\rm SL}(8,\mathbb{R})$, and the
gauge fields transform in the ${\bf 28}$ representation. In terms of
\eqn{spelec}, the generators are associated with $\tilde \Lambda$,
whereas \eqn{uspelec} is satisfied in view of the fact that both
matrices $\Lambda$ and $\Sigma$ are real. The ${\bf 56}$
representation of \Exc7 decomposes into the ${\bf 28}+\overline{\bf
28}$ representation of ${\rm SL}(8,\mathbb{R})$. The corresponding
Lagrangian is the one written down in \cite{deWitNic}.  To understand
the relation with the alternative, but inequivalent, Lagrangians
discussed below, we decompose the ${\bf 28}$ and $\overline {\bf 28}$
representation pertaining to the vector fields and their magnetic
duals, according to the ${\rm SL}(2,\mathbb{R})\times {\rm
SO}(1,1)\times {\rm SL}(6,\mathbb{R})$ subgroup,
\bea
\label{28-sl8}
{\bf 28}&\rightarrow&  ({\bf 2},{\bf 6})_{-1}+({\bf 1},{\bf
15})_{+1}+({\bf 1},{\bf 1})_{-3}\;,\nonumber \\
\overline{\bf 28}&\rightarrow&  ({\bf 2},\overline{\bf 6})_{+1}+({\bf
1},\overline{\bf 15})_{-1}+({\bf 1},{\bf 1})_{+3}\;.
\eea
There are no other symmetries of the Lagrangian beyond ${\rm
SL}(8,\mathbb{R})$. Of course, the combined field equations and
Bianchi identities have \Exc7 as a symmetry group, but this is true in 
general; it is only the symmetry group of the Lagrangian that can
differ, depending on the choice for ${\sf E}$.

\paragraph{The ${\rm E}_{6(6)}$-basis:
${\sf E}={\sf E}_{{\rm E}_{6(6)}}$} For a different choice of ${\sf
E}$ discussed below, there exists a larger group ${\rm G_e}$ which
contains the semisimple group ${\rm E}_{6(6)}\times {\rm
SO}(1,1)\subset {\rm E}_{7(7)}$ as a subgroup. The latter corresponds
to the generators $\tilde \Lambda$ in\eqn{spelec} and acts block
diagonally (the blocks $\Lambda,\,\Sigma$ are real). In addition,
there are 27 nilpotent generators which reside in both $\tilde
\Lambda$ and $\tilde \Sigma$ and extend the group to a non-semisimple
one. They transform in the $\overline{\bf 27}_{+2}$ representation of
${\rm E}_{6(6)}\times {\rm SO}(1,1)$ and give rise to the imaginary
parts of $\Lambda$ and $\Sigma$, although their combined contribution
to \eqn{uspelec} vanishes.  This ${\rm E}_{6(6)}$-basis is based on a
matrix ${\sf E}$ such that the Lagrangian coincides with the
Lagrangian that one obtains upon reduction to four dimensions of
five-dimensional ungauged maximally supersymmetric supergravity.

To understand the difference between this basis and the previous one,
let us consider the action of the common subgroup
${\rm SL}(2,\mathbb{R})\times {\rm SO}(1,1)\times {\rm SL}
(6,\mathbb{R})=\left({\rm E}_{6(6)}
\times {\rm SO}(1,1)\right)\cap {\rm SL}(8,\mathbb{R})$. With
respect to ${\rm E}_{6(6)}\times {\rm SO}(1,1)$
the ${\bf 56}$ of ${\rm E}_{7(7)}$ decomposes as:
\begin{eqnarray}
{\bf 56}&=&   \overline{\bf 27}_{-1} +{\bf
1}_{-3}+    {\bf 27}_{+1}+{\bf 1}_{+3} \,,
\label{56e6}
\end{eqnarray}
where the negative grading identifies the vector gauge fields and the
positive grading their magnetic duals. According to the
decomposition \eqn{spelec} the generators of ${\rm G_e}$ have
the form
\cite{AndDauFerrLle},
\begin{eqnarray}
\mbox{${\rm E}_{6(6)}$}&: &\left(\matrix{K_{27} &
0&\varnothing_{27}&0\cr 
0 & 0 & 0 & 0\cr \varnothing_{27} & 0 &-K_{27}^{\rm T}&0 \cr
0 & 0 & 0& 0 }\right)\;, \qquad
\mbox{${\rm SO}(1,1)$}:
\left(\matrix{-\bfone_{27} &
0&\varnothing_{27}&0\cr 0 & -3 & 0 & 0\cr \varnothing_{27} & 0
&\bfone_{27}&0 \cr 0 &
0 & 0& +3 }\right)\;,\nonumber\\[4mm]
\overline{\bf 27}_{+2}&:&\left(\matrix{\varnothing_{27} & \vec{t}
&\varnothing_{27}&0\cr 0 & 0 & 0 & 0\cr L_{27}& 0 &\varnothing_{27}&0
\cr 0 & 0 & -{\vec{t}}^{\,\rm T} & 0 }\right) \;,
\label{e66basis}
\end{eqnarray}
where $\vec{t}$ denotes a 27-dimensional vector of parameters, the
subscript 27 denotes $27\times 27$ matrices, and $K_{27}$ denotes the
generators of \Exc6 in the $\overline{\bf 27}$ representation;  the
symmetric matrix $L_{27}$ is expressed as
$L_{\Lambda\Sigma}=d_{\Lambda\Sigma\Gamma}\,t^\Gamma$,
where $d_{\Lambda\Sigma\Gamma}$ defines the cubic invariant in the
$\overline{\bf 27}$ representation of ${\rm E}_{6(6)}$. 
According to \eqn{56e6}, the 28 vector potentials $A^{AB}_\m$ now
transform in the $\overline{\bf 27}_{-1} + {\bf 1}_{-3}$
representation of ${\rm E}_{6(6)} \times {\rm SO}(1,1)$, which
decomposes with respect to the ${\rm SL}(2,\mathbb{R})\times {\rm
SO}(1,1)\times {\rm SL}(6,\mathbb{R})$ subgroup according to
\begin{eqnarray}
\overline{\bf 27}_{-1}+{\bf 1}_{-3}&\rightarrow &({\bf 2},{\bf
6})_{-1}+ ({\bf 1},\overline{\bf 15})_{-1}+({\bf 1},{\bf 1})_{-3}\,.
\label{sl8e6}
\end{eqnarray}
Comparing this result with the decomposition \eqn{28-sl8} shows
that the matrix ${\sf E}$ corresponds therefore to the duality
transformation that interchanges the ${\bf (1,15)}_{+1}$ gauge fields
of the ${\rm SL}(8,\mathbb{R})$-basis with the corresponding  dual
gauge fields belonging to the $({\bf 1},\overline{\bf 15})_{-1}$
representation. This duality transformation is not a symmetry of the
Lagrangian and the matrix ${\sf E}$ is thus not of the form
\eqn{spelec}. In the pseudoreal basis ${\sf E}$ is block diagonal
and ${\sf U}_{IJ}{}^{AB}$ acts as $i\bfone_{15}$ on the subspace
spanned by $({\bf 1},{\bf 15})_{+1}$ (the sign is irrelevant) and as
the identity on $({\bf 2},{\bf 6})_{-1}+ ({\bf 1},{\bf 1})_{-3}$. This
does not constitute an element of \Exc7. If that were the case the
matrix ${\sf U}$ should be an element of ${\rm SU}(8)$ acting in the
${\bf 28}$ representation. However, there is only one nontrivial
matrix ${\sf U}$ that is diagonal with eigenvalues 1 and/or $+i$,
namely, the matrix ${\rm diag}\;(\bfone_{12},i \bfone_{16})$. Hence we
conclude that ${\sf E}$ is indeed a nontrivial element of ${\rm
E}_{7(7)}\backslash {\rm Sp}(56;\mathbb{R})/{\rm GL}(28)$. Let us
denote this transformation by ${\sf E}_{{\rm E}_{6(6)}}$.

\paragraph{The ${\rm SL}(2,\mathbb{R})\times {\rm SO}(1,1)\times {\rm
SL}(6,\mathbb{R})$-basis: ${\sf E}={\sf E}_{{\rm
SL}(2,\mathbb{R})\times {\rm SO}(1,1)\times {\rm SL}(6,\mathbb{R})}$}
This basis is related to the two previous ones upon performing an
additional electric-magnetic duality transformation, this time also
interchanging the $({\bf 2},{\bf 6})_{-1}$ gauge fields of the ${\rm
SL}(8,\mathbb{R})$-basis with their dual gauge fields in the $({\bf
2},\overline{\bf 6})_{+1}$ representation, so that the gauge fields
decompose according to
\be
 ({\bf 2},\overline{\bf 6})_{+1}+ ({\bf
1},\overline{\bf 15})_{-1}+({\bf 1},{\bf 1})_{-3}\,.
\label{sl6sl2}
\ee
The semisimple invariance group ${\rm SL}(2,\mathbb{R})\times {\rm
SO}(1,1)\times {\rm SL}(6,\mathbb{R})$ is extended by 12 nilpotent
generators, which belong to the $({\bf 2},{\bf 6})_{+2}$
representation, so that ${\rm G}_{\rm e}$ is again a non-semisimple
group. To derive the above results is not difficult. One simply
decomposes the possible generators of \Exc7 in this duality rotated
basis in terms of ${\rm SL}(2,\mathbb{R})\times {\rm SO}(1,1)\times
{\rm SL}(6,\mathbb{R})$ and verifies which generators satisfy
\eqn{spelec} or \eqn{uspelec}. The new matrix ${\sf E}$ is equal to
${\sf E}^\prime\, {\sf E}_{{\rm E}_{6(6)}}$, where ${\sf E}^\prime$ is
again a block-diagonal matrix in which the submatrix ${\sf U}$ is
diagonal: ${\rm diag}\;(\bfone_{16}, i\bfone_{12})$, which, according
to the argument presented above, is not an element of ${\rm SU}(8)$,
so that the new matrix ${\sf E}$ belongs to another equivalence class.

\paragraph{The ${\rm SU^\star (8)}$-basis:
${\sf E}={\sf E}_{{\rm SU^\star (8)}}$} Another basis considered in
the literature \cite{Hull2}, is the one in
which ${\rm G_e}={\rm SU^\star (8)}$, the group of $8\times8$ matrices
that are real up to a symplectic matrix. This group is generated inside
${\rm E}_{7(7)}$ by the generators of the maximal compact subgroup
${\rm USp(8)}$ of ${\rm E}_{6(6)}$ and by the non-compact part of the
$\overline{\bf 27}_{+2}$ generators defined above. The vector
potentials transform in the {\bf 28} pseudoreal representation of
${\rm SU^\star (8)}$. The matrix ${\sf E}_{{\rm SU^\star (8)}}$ which
realizes the transformation from the ${\rm SL}(8,\mathbb{R})$-basis
to the ${\rm SU^\star (8)}$-basis can be expressed in the real
representation by means of the following transformation:
\begin{eqnarray}
{\sf E}_{{\rm SU^\star (8)}}&=&{\sf E}^{\prime\prime}\, {\sf E}_{{\rm
E}_{6(6)}}\in {\rm Sp(56,\mathbb{R})} \;,\nonumber\\[2mm]
 {\sf E}^{\prime\prime} &=&
\frac{1}{\sqrt{2}}\left(\matrix{\bfone_{27} & 0 
&-\bfone_{27}&0\cr 0 & 1 & 0 & 1\cr \bfone_{27} & 0 &\bfone_{27}&0 \cr
0 & -1 & 0 & 1 }\right)\,,
\label{e6sust}
\end{eqnarray}
where the matrix ${\sf E}^{\prime\prime}$ acts in the ${\rm
E}_{6(6)}$-basis of (\ref{e66basis}).
\section{The $T$-tensor}
\setcounter{equation}{0}
\label{T-tensor}
The gauging of supergravity is effected by switching on the gauge
coupling constant, after assigning the various fields to
representations of the gauge group embedded in G. Again we mainly
focus on the four-dimensional theory, where ${\rm G}={\rm E}_{7(7)}$,
but we will occasionally comment on other space-time dimensions. In
$d=4$ dimensions only the gauge fields themselves and the spinless
fields transform under the gauge group. In other dimensions, the
ungauged supergravity theory
has tensor gauge fields transforming in representations of the group
${\rm G}$. Coupling the vector gauge fields to tensor gauge fields
usually leads to a loss of tensor gauge invariance and may require
changes of the field representations. We already alluded to this
previously. However, for the analysis of this section, the tensor
gauge fields play no role and we concentrate on the vector gauge
fields and the scalar fields.

When switching on the charges, the abelian field strengths are changed
to nonabelian ones and derivatives of the scalars are covariantized
according to ({\it c.f.} \eqn{eq:gauged-cov-der})
\be
\label{gauge-bein}
\pa_\m \vv \to \pa_\m \vv - g A^{AB}_\m \,t_{AB} \vv \,,
\ee
where the gauge group generators $t_{AB}$ span a subalgebra of the Lie
algebra of  ${\rm G}_{\rm e}\subset {\rm E}_{7(7)}$ in the
${\bf 56}$ representation, whose
dimension is at most equal to the number of vector fields.
Introducing the gauging causes a loss of supersymmetry,
because the new terms in the Lagrangian yield new variations. The
leading variations are induced by the modification \eqn{gauge-bein} of
the Cartan-Maurer equations. This modification was already noted in
\eqn{CM-gauged} and takes the form
\bea
F_{\mu\nu}({\cal Q})_i{}^j & = &  -\ft43\,{\cal P}_{[\mu}{}^{\!jklm}
\, {\cal P}_{\nu]iklm} -g\,F_{\m\n}^{AB} \,{\cal Q}_{AB\,i}{}^j\,,
\nonumber \\
D_{[ \mu}^{~} {\cal P}_{\nu]}^{ijkl} &=& - \ft12 g\,F_{\m\n}^{AB}
\,{\cal P}_{AB}^{ijkl}
\,, \label{GECM-Q-P}
\eea
where
\be
\label{P-Q-gauged}
\vv^{-1} t_{AB} \vv = \pmatrix{{\cal Q}_{AB\, ij}{}^{\!mn}
&{\cal P}_{AB\, ijpq} \cr \noalign{\vglue 6mm}
{\cal P}_{AB}^{klmn} & {\cal Q}_{AB}{}^{\!kl}{}_{\!pq} \cr}\,.
\ee
These modifications are the result of the implicit dependence of
${\cal Q}_\m$ and ${\cal P}_\m$ on the vector potentials
$A_\m^{AB}$. This dependence can be expressed as follows,
\begin{eqnarray}
  \label{eq:extra-delta-P/Q}
{\cal Q}_\m = {\cal Q}_\m^{(0)} - g A_\m^{AB}\, {\cal Q}_{AB} \,,
\qquad
{\cal P}_\m = {\cal P}_\m^{(0)} - gA_\m^{AB}\, {\cal P}_{AB} \,.
\end{eqnarray}
The fact that the matrices $t_{AB}$ generate a subalgebra of the
algebra associated with ${\rm E}_{7(7)}$, implies that the quantities
${\cal Q}_{AB}$ and ${\cal P}_{AB}$ satisfy the constraints,
\bea
{\cal P}_{AB}^{ijkl} &=& \ft1{24}\,\varepsilon^{ijklmnpq}\, {\cal
 P}_{AB\,mnpq}\,, \nonumber \\
{\cal Q}_{AB\, ij}{}^{\!kl}&=& \d^{[k}_{[i}\, {\cal Q}_{AB\,
j]}{}^{\!l]}\,,
\eea
while ${\cal Q}_{ABi}{}^{\!j}$ is antihermitean and traceless.
It is straightforward to write down the explicit expressions for
${\cal Q}_{AB}$ and ${\cal P}_{AB}$,
\bea
{\cal Q}_{AB \,i}{}^{\!j} &=& \ft23 \Big[u_{ik}{}^{\!IJ}\, (\Delta_{AB}
u^{jk}{}_{\!IJ}) - v_{ikIJ}\, (\Delta_{AB} v^{jkIJ})  \Big]
 \,, \nonumber\\
{\cal P}_{AB}^{ijkl} &=&  v^{ijIJ}\,(\Delta_{AB} u^{kl}{}_{\!IJ})-
u^{ij}{}_{\!IJ} (\Delta_{AB} v^{klIJ})  \,.
\eea
where $\Delta_{AB}u$ and $\Delta_{AB}v$ indicate the change of
submatrices in $\vv$ induced by left multiplication with the generator
$t_{AB}$.

The presence of the order-$g$ modifications in the Cartan-Maurer
equations leads to new supersymmetry variations of
the gravitino kinetic terms and the Noether term. These variations are
proportional to the field strengths $F_{\m\n}^{AB}$, which must be
re-expressed in terms of the ${\rm SU}(8)$ field strengths
\eqn{su8-fs} in order to cancel against new supersymmetry variations
and terms in the Lagrangian. Hence the order-$g$
variations are proportional to the so-called $T$-tensor, which
decomposes into two reducible representations of ${\rm SU}(8)$ that
appear in the variations linear in the gravitino fields and linear in
the spin-$\ft12$ fields, respectively,
\bea
\label{T1-tensor}
T_i^{jkl} &=& \ft34{\cal Q}_{AB\,i}{}^j\, (u^{kl}{}_{\!AB} + v^{klAB})
\nn \\
&=& \ft12 \Big[u_{im}{}^{\!IJ}\, (\Delta_{AB}
u^{jm}{}_{\!IJ}) - v_{imIJ}\, (\Delta_{AB} v^{jmIJ})  \Big]
(u^{kl}{}_{\!AB} + v^{klAB})\,, \\
\label{T2-tensor}
T_{ijkl}^{mn} &=&\ft12 {\cal P}_{AB\,ijkl} \, (u^{mn}{}_{\!AB} +
v^{mnAB} ) \nn
\\
&=& \ft12\Big[ v_{ijIJ}\,(\Delta_{AB} u_{kl}{}^{\!IJ})-
u_{ij}{}^{\!IJ} (\Delta_{AB} v_{klIJ})\Big]
\, (u^{mn}{}_{\!AB} + v^{mnAB} )\,.
\eea
The $T$-tensor is thus a cubic product of the 56-bein $\vv$ and
depends in a nontrivial way on the embedding of the gauge group into
${\rm E}_{7(7)}$. It must satisfy a number of important
properties which are generic and apply to arbitrary spacetime
dimensions (of course, after switching to the corresponding G/H coset
space). These properties are discussed below.

First we observe that any variation of $\vv$  can be parametrized by
\be
\vv\to \vv \pmatrix{ 0&\overline\Sigma\cr\noalign{\vskip1mm}
\Sigma&0\cr} \,,
\ee
up to a (local) ${\rm SU}(8)$ transformation. Under this variation one
can easily show that the ${\rm SU}(8)$ tensors  ${\cal Q}_{AB}$ and
${\cal P}_{AB}$ combine into the ${\bf 133}$ representation of
${\rm E}_{7(7)}$. Likewise we can derive,
\bea
\label{T-variations}
\d T_i^{jkl} &=&  \Sigma^{jmnp}_{~} \,T^{kl}_{imnp} -\ft1{24}
\varepsilon^{jmnpqrst} \,\Sigma_{imnp}^{~} \,T^{kl}_{qrst} +
\Sigma^{klmn}_{~}\, T^j_{imn} \,, \nn\\
\d T_{ijkl}^{mn} &=& \ft43 \Sigma_{p[ijk}^{~}\, T^{pmn} _{l]} -\ft1{24}
\varepsilon_{ijklpqrs}\, \Sigma^{mntu}_{~}\, T^{pqrs}_{tu} \,.
\eea
This shows that the ${\rm SU}(8)$ covariant $T$-tensors constitute a
representation of ${\rm E}_{7(7)}$, corresponding to the
right multiplication of $\vv$. This property will play an
important role below.

The $T$-tensors are subject to quadratic equations, which are crucial
in establishing the supersymmetry at order $g^2$. Following
\cite{deWitNic} we first write $\hat \vv\, \hat\vv^{-1}=\bfone$ in
terms of the submatrices $u$ and $v$, and derive
\be
(u^{ij}{}_{\!CD} + v^{ijCD}) u_{ij}{}^{\!AB}  =
(u_{ij}{}^{\!CD} + v_{ijCD}) v^{ijAB} + \d_{CD}^{\!AB}\,.
\ee
Multiplication with ${\cal Q}_{CD}$ and ${\cal P}_{CD}$ and taking
suitable linear combinations yields, 
\bea
T^k_{lij} \, (u^{ij}{}_{\!AB}+v^{ijAB}) &=& -  T^{kij}_l\,
(u_{ij}{}^{\!AB}+v_{ijAB}) \,,\nonumber \\
T^{mn}_{ijkl} \, (u_{mn}{}^{\!AB}+v_{mnAB})
&=&\ft1{24}\varepsilon_{ijklpqrs} \; T^{pqrs}_{mn} \,
(u^{mn}{}_{\!AB}+v^{mnAB}) \, .
\eea
Contracting once more with ${\cal Q}_{AB}$ and ${\cal P}_{AB}$ yields
a number of identities quadratic in the $T$-tensors,
\bea
\label{quadratic}
T^k_{lij} \,T^{mij}_n - T^{kij}_l \, T_{nij}^m&=&0  \,,\nonumber \\ 
T^k_{lij} \,T^{ij}_{mnpq} +\ft1{24}\varepsilon_{mnpqrstu}\, T^{kij}_l
\, T^{rstu}_{ij}&=& 0  \,,\nonumber \\  
T^{vw}_{irst} \, T_{vw}^{jrst} - \ft18  \d_i^j\, T^{vw}_{rstu} \,
T_{vw}^{rstu}&=& 0 \,,\nonumber \\ 
T^{vw}_{ijkr} \,T_{vw}^{mnpr}
- \ft94 \d_{[\,i}^{[m} \, T^{vw}_{jk]rs} \,T_{vw}^{np]rs} 
+\ft1{16} \d{}_{\,i}^m{}_j^n{}_k^p \,T^{vw}_{rstu} \, T_{vw}^{rstu} 
  &=& 0 \,,
\eea
where in the last identity the antisymmetrization does not include
the indices $v,w$. 

The above considerations apply also to other dimensions. Observe that
the $T$-tensor transforms according to a tensor product of the
representation associated with the field strengths, in this case the
${\bf 56}$ of \Exc7, with the adjoint representation of G, in this
case the ${\bf 133}$ of \Exc7. Likewise, the $T$-tensor for $d=7$
belongs to the ${\bf 10}\times {\bf 24}$ representation of ${\rm
SL}(5)$, for $d=6$ it belongs to the ${\bf 16}\times {\bf 45}$
representation of ${\rm SO}(5,5)$, and for $d=5$ to the ${\bf
27}\times {\bf 78}$ representation of \Exc6. Obviously, these
representations are all reducible and, as we shall discuss in the next
section, a consistent gauging requires the $T$-tensor to take its
values in a smaller representation.

Before completing the analysis leading to a consistent gauging we
stress that all variations of the Lagrangian are expressed entirely in
terms of the $T$-tensor, as its variations is again proportional to
the same tensor. This includes the ${\rm SU}(8)$ covariant derivative
of the $T$-tensor, which follows directly from \eqn{T-variations} upon
the substitutions $\d\to D_\m$ and $\Sigma \to {\cal P}_\m$.  A viable
gauging requires that the $T$-tensor satisfies a number of rather
nontrivial identities, as we will discuss shortly, but the new terms
in the Lagrangian and transformation rules have a universal form,
irrespective of the gauge group (except from the term
\eqn{total-local}, which must be included depending on the nature of
the gauge group).  Let us first describe these new terms. First of
all, the order-$g$ variations from Cartan-Maurer equations are
cancelled by the variations of new masslike terms for the fermions,
\bea
\label{g-masses}
\lagr_{\rm masslike} & = & g\, e\Big\{ \ft12 \sqrt{2}\,  A_{1ij}
\overline{\psi}\,^i_\mu
\gamma^{\mu\nu} \psi^j_\nu + \ft{1}{6}  A_{2i}^{jkl} \,
\overline{\psi}^i_\mu \gamma^\mu \chi_{jkl}
\nonumber \\ & & {} \hspace{7mm}
+ A_3^{ijk,lmn}\,
\overline{\chi}_{ijk}\chi_{lmn} + {\rm h.c.}\Big \} \, ,
\eea
and by new terms in the supersymmetry transformations of the fermion
fields,
\bea
\label{g-variations}
\d_g\bar\psi^i_\m &=& -\sqrt{2} g \, A_1^{ij}\,\bar\e_j \gamma_\m
\,,\nn\\
\d_g\chi^{ijk}&=& - 2 g \,A_{2l}{}^{\!ijk}\,\bar\e^l\,.
\eea
Finally at order $g^2$, supersymmetry requires a potential for the
spinless fields,
\be
\label{g-potential}
P(\vv ) = g^2 \Big \{\ft{1}{24}  \vert A_{2i}{}^{\!jkl}\vert^2 -
\ft{1}{3} \vert A_1^{ij}\vert^2\Big\} \,.
\ee
In passing we note that the ${\rm SU}(8)$ covariant derivative of $A_1$
is proportional to $A_2$, 
\be
D_\m A_1^{ij} = \ft1{12} \sqrt{2} \, A_{2\,klm}^{(i}\, {\cal
P}_\m^{j)klm}  \,. 
\ee
A similar, but slightly more complicated result holds for the
derivative of $A_2$. 

The new terms in the Lagrangian and transformation rules have a form
that does not depend on the details of the gauging. Note that the
tensors $A_1^{ij}$, $A_{2i}{}^{\!jkl}$ and $A_3^{ijk,lmn}$ have
certain symmetry properties dictated by the way they appear in the
Lagrangian
\eqn{g-masses}: $A_1$ is symmetric in $(ij)$, $A_2$
is fully antisymmetric in $[jkl]$ and $A_3$ is antisymmetric in
$[ijk]$ as well as in $[lmn]$ and symmetric under the interchange
$[ijk]\leftrightarrow[lmn]$. Therefore these tensors transform
under ${\rm SU}(8)$ according to the representations
\bea
\label{A:tosu8}
A_1&:& {\bf 36}+ \overline{\bf 36}\,,\nn \\
A_2&:& {\bf 28}+ \overline{\bf 28}+ {\bf 420}+ \overline{\bf 420}\,,
\nn\\ 
A_3&:& {\bf 420}+ \overline{\bf 420}+{\bf 1176}+  \overline{\bf 1176}
\,.  
\eea

This analysis can be repeated for each of the maximal
supergravities. In table~\ref{mass-repr} we indicate all possible
masslike terms for the fermions by indicating the irreducible
representations of ${\rm H}_{\rm R}$ to which they belong. These
results are relevant for deducing the constraints on the
$T$-tensor in the next section. We observe here
that only in $d=7$ dimensions the matrices $A_1$ and $A_3$ are 
antisymmetric, due to the property of pseudo Majorana spinors. 

\begin{table}[tb]
\begin{center}
\begin{tabular}{l l l l l }\hline
$d$ &${\rm H}_{\rm R}$& $A_1$ & $A_2$ & $A_3$  \\   \hline
7   & USp(4)   & ${\bf 1}+{\bf 5}$ & ${\bf 5}+ {\bf 10}+{\bf 14}+ {\bf
35}$ 
   & ${\bf 1}+{\bf 5}+{\bf 14} +{\bf 30}+{\bf 35}+{\bf 35}^\prime$
\\[.5mm] 
6  & ${\rm USp}(4)\!\times\!{\rm USp}(4)$ & $({\bf 4},{\bf 4})$ &
  $\!({\bf 4},{\bf 4})+ ({\bf 4},{\bf 4})$   &
  $({\bf 4},{\bf 4})
  + ({\bf 4},{\bf 16}) +({\bf 16},{\bf 4})$\\
~ & ~ & ~ & $+ ({\bf 4},{\bf 16}) +({\bf 16},{\bf 4})$
 & $+ ({\bf 16},{\bf 16})$ \\[.5mm]
5   & USp(8) & ${\bf 36}$ & ${\bf 27}+ {\bf 42}+ {\bf 315}$ & ${\bf 1}
  +{\bf 27}+{\bf 36}+{\bf 308}$  \\
~&~&~&~&$+{\bf 315}+{\bf 792}+{\bf 825}$  \\[.5mm]
4   & SU(8)  & ${\bf 36}+\overline{\bf 36}$ & ${\bf 28}
  + \overline{\bf 28}+ {\bf 420}+ \overline{\bf 420}$ &
  ${\bf 420}+\overline{\bf 420}+ {\bf 1176} +\overline{\bf 1176}$
\\[.5mm]
3   & SO(16) & ${\bf 1}+ {\bf 135}$ & ${\bf 128}+ \overline{\bf 1920}$
  & ${\bf 1}+ {\bf 1820}+\overline{\bf 6435}$ \\ \hline
\end{tabular}
\end{center}
\caption{\small
Possible fermion mass terms for maximal supergravities in various
dimensions assigned to irreducible R-symmetry representations. Note
that in $d=7$ dimensions the tensors $A_1$ and $A_3$ are antisymmetric
in the fermion indices.  }\label{mass-repr}
\end{table}

Because covariant variations of the $T$-tensor are again proportional
to the $T$-tensor, and the potential is ${\rm SU}(8)$ invariant,
variations of the potential are quadratic in the
$T$-tensor. Stationary points of the potential are subject to the
condition that $\Omega^{ijkl}$ must be anti-selfdual, where the tensor
$\Omega$ is defined by \cite{deWitNic-par},
\be
{\Omega}^{ijkl} = \ft34 A_{2\,m}^{n[ij}\,A_{2\,n}^{kl]m} - A_1^{m[i}\,
A_{2\,m}^{jkl]} \,. 
\ee
Expanding the potential about a stationary point leads to a mass term,
which is again quadratic in the $T$-tensor. We refer to the explicit
expressions given in \cite{deWitNic-par}. All this is completely
generic and similar formulae can be derived for maximal supergravity
in any spacetime dimension.

The mass term for the vector fields is generated by the gauge
covariantizations in the scalar kinetic term. Making use of
\eqn{eq:extra-delta-P/Q}, one arrives immediately at the following
relation 
\be
\lagr= -\ft1{12} e \,\vert{\cal P}^{ijkl}_\mu\vert^2 \longrightarrow
-\ft 1{12} e g^2\,A^{AB}_\mu\,A^{\m\,CD}\,
\mathcal{P}_{AB\,ijkl}\,\mathcal{P}_{CD}^{ijkl}\,.   
\label{vecmass} 
\ee
The physical mass is, however, given in terms of the square of the
$T$-tensor as one notes after combining the mass term with the kinetic
terms for the vector fields,
\bea
\lagr &=&  -\ft18 e\,\left[
F_{\m\n}^{AB}[(u+v)^{-1}]^{AB}{}_{\!ij}\right]\, 
\left[ F^{\m\n\, CD}[(u+v)^{-1}]_{CD}{}^{\!kl}\right]\; \d^{ij}_{kl}
\nonumber \\ 
&& -\ft13 e\, g^2 \,\left[A_{\m}^{AB}[(u+v)^{-1}]^{AB}{}_{\!ij}
\right] \,\left[ A^{\m\,
CD}[(u+v)^{-1}]_{CD}{}^{\!kl}\right] \; T^{ij}_{mnpq}\,T_{kl}^{mnpq}
\,, 
\eea
where we suppressed the terms proportional to
$\varepsilon^{\m\n\rho\sigma} F^{AB}_{\m\n}F^{CD}_{\rho\sigma}$ as
well as the terms~\eqn{total-local}.

\section{Group-theoretic analysis}
\setcounter{equation}{0}

The three ${\rm SU}(8)$ covariant tensors, $A_1$, $A_2$ and $A_3$,
which depend only on the spinless fields, must be linearly related to
the $T$-tensor, because they were introduced for the purpose of
cancelling the variations proportional to the $T$-tensors. To see how
this can be the case, let us analyze the ${\rm SU}(8)$ content of the
$T$-tensor. As we mentioned
already, the $T$-tensor is cubic in the 56-bein, and as such it
constitutes a tensor that transforms under ${\rm
E}_{7(7)}$. The transformation properties were given in
\eqn{T-variations}, where we made use of the fact that the $T$-tensor
consists of a product of the fundamental times the adjoint
representation of ${\rm E}_{7(7)}$. Hence the $T$-tensor comprises the
representations,
\be
\label{56x133}
{\bf 56}\times{\bf 133} = {\bf 56} + {\bf 912} + {\bf 6480} \,.
\ee
The representations on the right-hand side can be decomposed under
the action of ${\rm SU}(8)$,
\bea
\label{T:e7tosu8}
{\bf 56}&=& {\bf 28}+ \overline{\bf 28}\,, \nn\\
{\bf 912}&=& {\bf 36}+\overline{\bf 36} + {\bf 420}
+ \overline{\bf 420}\,, \nn \\
{\bf 6480}&=& {\bf 28}+ \overline{\bf 28} +{\bf 420}+
\overline{\bf  420} +{\bf 1280}+ \overline{\bf 1280}
+{\bf 1512}+ \overline{\bf 1512} \,.
\eea
These representations should correspond to the ${\rm SU}(8)$
representations to which the tensors $A_1$-$A_3$ (and their complex
conjugates) belong. However, there is a mismatch between
\eqn{T:e7tosu8} and \eqn{A:tosu8}, which shows that the $T$-tensor is
constrained. In view of \eqn{T-variations} this constraint should
amount to suppressing complete representations of ${\rm E}_{7(7)}$ in
order that its variations and derivatives remain consistent. Therefore
the conclusion is that the $T$-tensor cannot contain the ${\bf 6480}$
representation of ${\rm E}_{7(7)}$, so that it consists at most of the
${\bf 28}+{\bf 36}+{\bf 420}$ representation of ${\rm SU}(8)$ (and its
complex conjugate). This implies that $A_3$ is not an independent
tensor and can be expressed in terms of $A_2$ and that the $T$-tensor
is decomposable into $A_1$ and $A_2$.  Indeed this was found by
explicit calculation, which reveals the relations
\bea
T_i^{jkl} &=& - \ft{3}{4} A_{2i}{}^{\!jkl} + \ft{3}{2}
A_1^{j[k}\,\delta_i^{l]} \,,\nn \\
T^{mn}_{ijkl} &=& - \ft 43 \d^{[m}_{[i} T^{n]}_{jkl]}\,,\nn\\
A_3^{ijk,lmn}&=&- \ft1{108}\sqrt{2} \, \varepsilon^{ijkpqr[lm}
T_{pqr}^{n]} \,.
\eea
Observe that the first equation implies that the ${\bf 28}$
representation is also suppressed, because the combination of
$T_i^{ikl}= 0$ with this equation implies that 
\be
T_i^{[ijk]}=0\,.
\ee
Hence the $T$-tensor transforms under ${\rm E}_{7(7)}$ according to
the ${\bf 912}$ representation which decomposes into the ${\bf 36}$
and ${\bf 420}$ representations of ${\rm SU}(8)$ and their complex
conjugates residing in the tensors $A_1$ and $A_2$, respectively,
\be
A_1^{ij} = \ft4{21} T_k^{ikj}\,,\qquad A_{2i}^{jkl} = -\ft 43
T_i^{[jkl]}\,.
\ee
The fact that the $T$-tensor is restricted to a particular
representation of ${\rm E}_{7(7)}$ ensures that the identities
\eqn{quadratic} quadratic in the $T$-tensor suffice to cancel the
order $g^2$ variations in the Lagrangian. Therefore the condition that
the $T$-tensor belongs to the ${\bf 912}$ representation is a
sufficient criterion for establishing the viability of a given
gauging.

We concentrated on the $d=4$ theory, but many of the above features
are generic and apply in other dimensions, where the complications
related to electric-magnetic duality are absent. For instance, we show
the representations of the unrestricted $T$-tensors for $3\leq d\leq
7$ spacetime dimensions in table~\ref{T-tensor-repr}.\footnote{
The $d=3$ theory has initially no vector fields, but those can be
included by adding Chern-Simons terms. These gauge fields are used to
gauge some of the ${\rm E}_{8(8)}$ isometries
\cite{NicSam}. }
\begin{table}
\begin{center}
\begin{tabular}{l l l  }\hline
$d$ &${\rm G}$& $T$  \\   \hline
7   & ${\rm SL}(5)$  & ${\bf 10}\times {\bf 24}= {\bf 10}+{\bf 15}+
 {\bf 40}+ {\bf 175}$  \\[.5mm]
6  & ${\rm SO}(5,5)$ & ${\bf 16}\times{\bf 45} =
  {\bf 16}+ {\bf 144} + {\bf 560}$ \\
5   & ${\rm E}_{6(6)}$ & ${\bf 27}\times{\bf 78} = {\bf 27} 
+ {\bf 351} + 
 {\bf 1728}$  \\[.5mm]  
4   & ${\rm E}_{7(7)}$  & ${\bf 56}\times{\bf 133} = {\bf 56} + {\bf
912} + 
{\bf 6480}$   \\[.5mm]
3   & ${\rm E}_{8(8)}$ & ${\bf 248}\times{\bf 248} = {\bf 1}
+ {\bf 248} + {\bf 3875} +{\bf 27000} +  {\bf 30380}$ \\ \hline
\end{tabular}
\end{center}
\caption{\small
Decomposition of the $T$-tensor in various
dimensions for maximal supergravities in terms of irreducible
representations of G.
}\label{T-tensor-repr}
\end{table}

In $d=3$ dimensions the ${\bf 248}+{\bf 30380}$ representation should
be suppressed as it is antisymmetric and cannot appear in the
Chern-Simons coupling. Just as above, comparing the R-symmetry
representations contained in the G-representations listed in
table~\ref{T-tensor-repr}, to the R-symmetry representations for the
masslike terms listed in table~\ref{mass-repr}, shows that the
following representations cannot appear: for $d=7$ the representation
${\bf 175}$, for $d=6$ the ${\bf 560}$ representation, for $d=5$ the
${\bf 1728}$ representation, for $d=4$ the ${\bf 6480}$ representation
and for $d=3$ the ${\bf 27000}$ representation. Following the
arguments presented above, we establish the following representation
content for the $T$-tensors and their branching into R-symmetry
representations:
\be
\begin{array}{rclcl}
\label{embedd-d}
d&\!\!=\!\!&7 &:& {\bf 15}\to {\bf 1}+{\bf 14} \;, \\
d&\!\!=\!\!&6 &:& {\bf 144}\to ({\bf 4},{\bf 4})
+({\bf 4},{\bf 16})+({\bf 16},{\bf 4}) \;, \\
d&\!\!=\!\!&5 &:& {\bf 351}\to {\bf 36}+ {\bf 315}\;, \\
d&\!\!=\!\!&4 &:& {\bf 912}\to {\bf 36}+\overline{\bf 36} + {\bf 420}
+ \overline{\bf 420}  \;, \\
d&\!\!=\!\!&3 &:& {\bf 1}+ {\bf 3875} \to  {\bf 1}+ {\bf 135}+{\bf
1820} + \overline{\bf 1920} \;.
\end{array}
\ee
These constraints on the $T$-tensor must be satisfied for any
gauging. All the corresponding R-symmetry representations must appear
in the Lagrangian with the right multiplicity. It turns out that, with
the exception of $d=3$, all representations are covered by the tensors
$A_1$ and $A_2$, so that the tensor $A_3$ is not independent. For
$d=3$, the tensor $A_3$ contains the  ${\bf 1820}$ representation of
${\rm SO}(16)$, which is not present in $A_1$ and $A_2$. Therefore
$A_3$ is independent in this case.

It is possible to rephrase some of the above in the following way.  A
gauging is characterized by a real {\it embedding matrix}
$\Theta_{M}{}^{\alpha}$ which defines how the gauge group is embedded
into ${\rm E}_{7(7)}$. Here the indices $\a$ and $M$ belong to the
adjoint and the fundamental representation of ${\rm E}_{7(7)}$,
respectively. Hence, $\a=1,\ldots, 133$, and $M=1,\ldots,56$. The
gauge group generators can now be labeled by indices belonging to the
fundamental representation,
\be
 t_M= \Theta_M{}^\a\, t_\a\,,
\ee
where the $t_\a$ are the 133 generators of ${\rm E}_{7(7)}$ in an
arbitrary representation. The fact that the $t_M$ generate a group,
implies that the embedding matrix satisfies the condition,
\be
\label{gauge-gen}
\Theta_M{}^\a\,\Theta_N{}^\b \,f_{\a\b}{}^{\g}= f_{MN}{}^P\,
\Theta_P{}^\g\,,
\ee
where the $f_{\a\b}{}^\g$ and $f_{MN}{}^P$ are the structure constants
of ${\rm E}_{7(7)}$ and the gauge group, respectively. This condition
implies that the embedding matrix is invariant under the gauge
group. In principle one is dealing with 28 electric and 28 magnetic
charges, which transform according to the ${\bf 56}$
representation. Only the electric charges can couple locally and the
subset of the corresponding generators that are involved in this
gauging were previously denoted by $t_{AB}$. As previously stated,
their embedding in the ${\bf 56}$ is constrained by the condition
(\ref{uspelec}), namely that they should be contained in the global
symmetry group ${\rm G_e}$ of the Lagrangian or, in other words, that
they correspond to {\it electric} charges.  The embedding matrix is in
fact directly related to the $T$-tensor, which transforms under ${\rm
E}_{7(7)}$ according to the tensor product ${\bf 56}\times {\bf
133}$. Namely, by making a field-dependent ${\rm E}_{7(7)}$
transformation the matrix $\vv$ can be reduced to the identity, so
that the $T$-tensor is field independent and equal to the generators
$t_{AB}$ appropriately contracted by ${\sf E}$. Recalling that the
indices of the ${\bf 56}$ refer to antisymmetric index pairs that are
denoted by lower or upper indices, we find for $M=[IJ]$ with lower
indices,
\be
T_M{}^\a\,t_\a\Big\vert_0 =  \Theta_M{}^\a\,t_\a
=\ft12  ({\sf U}_{IJ}{}^{\!AB}+ {\sf V}_{IJAB})\,
t_{AB} \,,
\label{TTheta0}
\ee
where here and henceforth the $t_\a$ are the ${\rm E}_{7(7)}$
generators in the fundamental representation. The complex conjugate of
this relation refers to the embedding matrix with $M$ referring to the
upper indices $[IJ]$. The $T$-tensor is obtained by transforming
$T\vert_0$ under ${\rm E}_{7(7)}$ with a field dependent
transformation defined by $\vv(x)$:
\be
T_M{}^\a[\Theta]\,t_\a = \vv^{-1\,N}_{\;M}\,\Theta_N{}^\a\,
\vv^{-1} t_\alpha \vv \;.
\label{TTheta}
\ee

As a final condition, which was explained above, the $T$-tensor is
restricted to be contained in the ${\bf 912}$
representation. Therefore the embedding matrix has to belong to that
representation. Defining appropriate projection operators, this
condition can be expressed as follows,
\begin{eqnarray}
\label{constraint-912}
\mathbb{P}_{(912)M}{}^{\!\alpha\, N}{}_{\!\beta}\,\Theta_N{}^\beta &=&
\Theta_M{}^\alpha\,,
\end{eqnarray}
where $\mathbb{P}_{(912)}$ is the projection operator on the ${\bf
912}$ representation in the product space \eqn{56x133}. As we explain
in detail in the appendix, this projector may be explicitly written in
terms of the generators $t_\alpha$ as
\begin{eqnarray}
\label{E7-projectors}
\mathbb{P}_{(912)M}{}^{\!\alpha\, N}{}_{\!\beta}
&=&-\ft{12}{7}\,(t^\alpha)_K{}^{\!N}\,(t_\beta)_M{}^{\!K} + \ft4{7}
(t^\alpha)_M{}^{\!K}\,(t_\beta)_K{}^{\!N}
+\ft{1}{7}\,\delta_M{}^{\!N}\delta^\alpha{}_{\!\beta}\,, \nn\\
\mathbb{P}_{(6480)M}{}^{\!\alpha\, N}{}_{\!\beta}
&=& \ft{12}{7}\,(t^\alpha)_K{}^{\!N}\,(t_\beta)_M{}^{\!K}
-\ft{132}{133} (t^\alpha)_M{}^{\!K}\,(t_\beta)_K{}^{\!N}
+\ft{6}{7}\,\delta_M{}^{\!N} \delta^\alpha{}_{\!\beta}\,, \nn\\
\mathbb{P}_{(56)M}{}^{\!\alpha\, N}{}_{\!\beta}
&=&\ft{8}{19}\,(t^\alpha)_M{}^{\!K}\,(t_\beta)_K{}^{\!N}\,,
\end{eqnarray}
where indices $\a,\b$ are raised and lowered with the invariant metric
$\eta_{\alpha \beta}={\rm Tr}(t_\alpha t_\beta)$. 

Similarly, in five dimensions the projectors on the ${\bf 351}$, ${\bf
1728}$ and the ${\bf 27}$ representations appearing in the branching
rules shown in table~\ref{T-tensor-repr} are obtained from
(\ref{projectors}) 
\begin{eqnarray}
\label{E6-projectors} \mathbb{P}_{(351)\Lambda}{}^{\!a\,
\Sigma}{}_{\!b}
&=&-\ft{6}{5}\,(t^a)_\Gamma{}^{\!\Sigma}\,(t_b)_\Lambda{}^{\!\Gamma} +
\ft3{10} (t^a)_\Lambda{}^{\!\Gamma}\,(t_b)_\Gamma{}^{\!\Sigma}
+\ft{1}{5}\,\delta_\Lambda{}^{\!\Sigma}\delta^a{}_{\!b}\,, \nn\\
\mathbb{P}_{(1728)\Lambda}{}^{\!a\, \Sigma}{}_{\!b} &=&
\ft{6}{5}\,(t^a)_\Gamma{}^{\!\Sigma}\,(t_b)_\Lambda{}^{\!\Gamma}
-\ft{42}{65} 
(t^a)_\Lambda{}^{\!\Gamma}\,(t_b)_\Gamma{}^{\!\Sigma}
+\ft{4}{5}\,\delta_\Lambda{}^{\!\Sigma} \delta^a{}_{\!b}\,, \nn\\
\mathbb{P}_{(27)\Lambda}{}^{\!a\, \Sigma}{}_{\!b}
&=&\ft{9}{26}\,(t^a)_\Lambda{}^{\!\Gamma}\,(t_b)_\Gamma{}^{\!\Sigma}\,,
\end{eqnarray}
where $\Lambda,\,\Gamma,\,\Sigma=1,\dots ,27$ and $a,\,b=1,\dots ,78$
label the basis elements in the fundamental and the adjoint
representation of \Exc6, respectively. The projector
$\mathbb{P}_{(351)}$ on the space of the embedding matrices
$\Theta_{\Lambda}{}^{a}$ will be used to implement the supersymmetry
restriction on the $T$-tensor and thus to define the allowed
gaugings.

As we shall see in the sequel, when we consider the four-dimensional
gaugings in the ${\rm E}_{6(6)}$-basis, the ${\bf 351}$
representation can be obtained from the branching of the ${\bf 912}$
of ${\rm E}_{7(7)}$ with respect to the \Exc6 subgroup and the
corresponding projector in the space of the embedding matrices can be
derived as a suitable restriction of $\mathbb{P}_{(912)}$.

The above projection operators are used in the computer-aided analysis
of the $T$-tensor that we will make use of in later sections.

%
\section{Scherk-Schwarz reductions}
\label{Scherk-Schwarz}

\setcounter{equation}{0}
In this section we summarize a number of features related to the
modified dimensional reduction scheme proposed by Scherk and Schwarz
\cite{ScherkSchwarz}. This scheme applies to a theory in higher
dimensions with a rigid internal symmetry group G. It consists of an
ordinary dimensional reduction on a hypertorus, where an extra
dependence on the torus coordinates is introduced by applying a
finite, uniform G-transformation that depends nontrivially on the
these coordinates. Subsequently one retains only the lowest Fourier
components. Because of the invariance the internal symmetry
transformation cancels out up to those terms in the Lagrangian that
contain derivatives with respect to the torus coordinates. As one can
easily verify, this reduction defines a consistent truncation of the
higher-dimensional theory, which thus corresponds to a deformation of
the lower-dimensional theory that one obtains by ordinary dimensional
reduction. The dependence of the internal symmetry transformation
deserves some comment. When compactifying on a hypertorus $T^n$ with
coordinates $y^m$, one specifies an $n$-dimensional abelian subgroup
of the internal symmetry group G, with generators $t_m$ so that the
group element $\hat g(y)\in{\rm G}$ equals $\hat g(y) = \exp[g
\,y^m\,t_m]$. For convenience we have introduced a coupling constant
$g$. The deformation of the lower-dimensional theory is governed by
the Lie-algebra valued quantities $\hat g^{-1}\pa_m \hat g$ which are
obviously $y$-independent and equal to the matrices $g\, t_m$. The
deformed theory is always related to a possible gauging of the theory,
with the parameter $g$ playing the role of the gauge coupling
constant, because the graviphotons couple to the $y$-dependent
quantities. The $t_m$ are the charges that couple to the graviphotons
according to the description given in earlier sections. The other
gauge fields that participate in the gauging are the gauge fields that
already exist in higher dimensions (or possibly, tensor fields that
give rise to vectors in lower dimensions). Because these fields
transform under the internal symmetry group in higher dimensions, they
will generically be charged with respect to the graviphotons. For
maximal supergravity, the gaugings define the only known
supersymmetric deformations. In this paper we start from the other
end; we analyze possible gaugings in lower dimensions and identify
some of them afterwards with the result of a Scherk-Schwarz
reduction. The inequivalent Scherk-Schwarz gaugings correspond to the
different conjugacy classes of $\hat g$.

In order to properly identify a Scherk-Schwarz gauging we consider a
generic Lagrangian of gravity in $d+n$ spacetime dimensions, coupled
to scalars, vectors and higher-rank antisymmetric gauge fields. The
Lagrangian depends only on Newton's constant and there are no other
dimensionfull coupling constants. In addition, the Lagrangian is
invariant under a group G and the scalar fields are assumed to
parametrize a homogeneous target space. Hence we consider a typical
Lagrangian with an Einstein-Hilbert term, scalar fields and a $k$-rank
gauge field, with possible interactions,
\be
\lagr_{d+n} = -{1\over 2\,\kappa_{d+n}^2} E \left[  R +
{\cal P}_M^{\,2} +
{(k+1)^2\over k!} \Big(\pa_{[M_1} A_{M_2\cdots M_{k+1}]}\Big)^2
+\cdots\right] \,,
\ee
as well as a single coupling constant $\kappa^2_{d+n}$. The kinetic
term for the tensor gauge field may be modified by terms depending on
the scalar fields, but this dependence has been suppressed. For
convenience of notation, only in this section the indices $M, N\dots$
and $A, B\dots$ denote $(d+n)$-dimensional world and target indices;
the world and target indices of the torus are $m, n\dots$ and $a,
b\dots$ while those of the $d$-dimensional space-time are $\mu,
\nu\dots$ and $\alpha, \beta\dots$.  The vielbein field is denoted by
$E_M{}^A$ and $E =\det (E_M{}^A)$. We assume that the Lagrangian
remains unchanged under a simultaneous rescaling of the coupling
constant and the fields,
\be
\label{scale-d+n}
\begin{array}{rcl}
E_M{}^A &\to& {\rm e}^{-\a} E_M{}^A\,,\\
{\cal P}_M&\to& {\cal P}_M \,,
\end{array}
\qquad
\begin{array}{rcl}
A_{M_1\cdots M_k} &\to& {\rm e}^{-k\a} A_{M_1\cdots M_k} \,,\\
\kappa^2_{d+n} &\to& {\rm e}^{(2-d-n)\a}\, \kappa_{d+n}^2\,.
\end{array}
\ee
Note that the tangent space tensors corresponding to ${\cal P}_M$ and
the field strengths (which may include modifications with scalar
fields), $F_{M_1\cdots M_{k+1}}= (k+1)\,\pa_{[M_1}A_{M_2\cdots
M_{k+1}]}$, scale with the same factor $\exp[\a]$. The composite
connection field ${\cal Q}_M$, which does not appear explicitly in the
above formula, scales precisely as ${\cal P}_M$.  All supergravity
Lagrangians that follow from 11-dimensional supergravity by standard
dimensional reduction, are of this type.

Under standard compactification on a torus $T^n$ the Lagrangian that
describes the massless modes in $d$ spacetime dimensions remains
invariant under the symmetry group G of the original
$(d+n)$-dimensional theory, and under ${\rm GL}(n)$. While the
transformations under the ${\rm SL}(n)$ subgroup are obvious from the
index structure of the various fields after dimensional reduction,
this is not so for the scale transformation. These scale
transformations will be denoted by ${\rm SO}(1,1)$ in the next
sections. Although the symmetries are not preserved by the modified
dimensional reduction, one can still classify the various fields with
respect to these transformations. They originate from a combined
uniform rescaling of the torus coordinates and \eqn{scale-d+n}, such
that the coupling constant $\kappa^2_d$ in $d$ dimensions remains
constant. The latter coupling constant is inversely proportional to
the torus volume.

Upon dimensional reduction the vielbein field decomposes
into the $d$-dimensional vielbein field $e_\m{}^\a$, $n$ -photon
fields $B^{\,m}_\m$, a graviscalar $\phi$ and an internal vielbein
field $\hat e_m{}^a$ with $\det[\hat e]=1$. The latter parametrizes a
${\rm SL}(n)/{\rm SO}(n)$ target space. In order to re-obtain
the Einstein-Hilbert Lagrangian in $d$ dimensions, one performs a
Weyl rescaling of the vielbein $e_\m{}^\a\to \exp[-n\phi/(d-2)]
\,e_\m{}^\a$. The tensor field decomposes into tensors that transform
irreducibly with respect to both the $d$-dimensional Lorentz
transformations and ${\rm SL}(n)$. The tensor fields are in general
not redefined by Weyl rescalings as this would affect the form of
their gauge
transformations. These steps are rather standard and we refrain from
giving further details. After they have been carried out, the fields
$e_\m{}^\a$, $\hat e_m{}^a$, ${\cal P}_\m$ and ${\cal Q}_\m$ remain
unaffected by the scale transformation, while the remaining fields
scale as follows,
\bea
\label{torus-scale}
{\rm e}^\phi  &\to&  {\rm e}^{(d-2)\a/n} \, {\rm e}^\phi\,,\nn\\
B^m_\m        &\to&  {\rm e}^{-[(d-2)/n +1]\a} \, B^m_\m \,,\nn\\
A_{\m_1\cdots\m_k}&\to&  {\rm e}^{-k\,\a} \,A_{\m_1\cdots\m_k} \,,\nn\\
A_{m_1\cdots m_p\,\m_{p+1}\cdots\m_k}&\to&  {\rm e}^{[(d-2)p/n
+p-k]\a} \,
A_{m_1\cdots m_p\,\m_{p+1}\cdots\m_k} \,,\nn\\
A_{m_1\cdots m_k}&\to&  {\rm e}^{k(d-2)\a/n}
\,A_{m_1\cdots m_k} \,.
\eea

In the Lagrangian the invariance under the scale transformations
reflects itself in certain exponential factors of the graviscalar. In
standard dimensional reduction, derivatives with respect to the torus
coordinates vanish and the scale invariance is exact.  In the
Scherk-Schwarz scheme these derivatives no longer vanish and give rise
to mass terms and a potential which break this invariance. The same
phenomenon is encountered when one retains the massive modes in
standard dimensional reduction where the Kaluza-Klein masses and
charges break the scale invariance. These masses and charges are
inversely proportional to the periodicity lengths associated with the
torus. However, when one combines the scale transformations with a
simultaneous, uniform rescaling of the torus (and thus of the
Kaluza-Klein masses and charges) then the Lagrangian remains unchanged.
For the Scherk-Schwarz reduction one
has the same situation. The group element $\hat g(y)$ is not
invariant under a uniform scaling of the $y^m$, but we can
simultaneously scale the coupling constant $g$ such that $g\,y^m$
remains invariant. Consequently, the Lagrangian is not invariant under
the scale transformations for fixed $g$, but it remains unchanged
provided we scale $g$ as well. In this way, the lack of scale
invariance is precisely characterized by the corresponding power of
$g$.

Suppressing the higher Fourier modes and
assigning the following  scale transformation to the coupling constant
$g$,
\be
g\to {\rm e}^{[(d-2)/n+1]\,\a }\,g\,,
\ee
the Lagrangian of the reduced theory remains unchanged under the
combined scale
transformations. Specifically, the following quantities which
involve derivatives $\pa_m$, scale consistently under uniform scale
transformations of the torus coordinates, {\it i.e.},
\bea
{\cal P}_m    &\to& {\rm e}^{[(d-2)/n+1]\a}\,{\cal P}_m\,,  \nn\\
{\cal Q}_m    &\to& {\rm e}^{[(d-2)/n+1]\a}\,{\cal Q}_m\,,  \nn\\
F_{m_1\cdots n_p\,\m_{p+1}\cdots\m_{k+1}}&\to&
  {\rm e}^{[(d-2)p/n +p-k]\a} \,
F_{m_1\cdots n_p\,\m_{p+1}\cdots\m_{k+1}} \,,
\eea
by virtue of the scale transformation of the coupling constant. Note
that the field strengths will in general be modified by certain
functions of the scalar fields (originating from the original
$(d+n)$-dimensional theory), but those modifications do not affect the
scaling behavior. With the above transformation rules, the 
(non-negative) potential,
\be
\label{SS-potential}
P = {e\over 2\,\kappa^2_d} \left[ \Big(
{\rm e}^{-[1+ n/(d-2)]\phi} \, {\cal P}_m\Big)^2  +{1\over k!}
\Big({\rm e}^{- [k+1+ n/(d-2)]\phi} \, F_{m_1\cdots m_{k+1}}
\Big)^2 \right] \;,
\ee
remains unaffected. For finite values of $\phi$ these potentials can
only have stationary points when ${\cal P}_m$ and $F_{m_1\cdots
m_{k+1}}$ vanish. These stationary points have zero cosmological
constant. When $\phi$ tends to infinity, the potential will vanish as
well. We return to the the properties of the Lagrangian at the end of
this section. 

The terms in parentheses can be interpreted as components of the
$T$-tensors, multiplied with the coupling constant $g$. Therefore it
follows that the $T$-tensors scale according to
\be
\label{so-weight}
T\to {\rm e}^{-[(d-2)/n +1]\,\a } \, T\,,
\ee
so that $g\,T$ remains invariant. Likewise also $g\,B^m{}_{\!\!\m}$
remains invariant. We can be a little more precise here, by 
also considering the fermions and following what happens to the
transformation rules under the reduction. After proper Weyl rescalings
and rediagonalization, one then concludes that the Scherk-Schwarz
$T$-tensors decompose as follows,
\bea
g\,A_1 &=& {\rm e}^{-[1+n/(d-2)]\phi} \Big[ {\cal Q}_m \oplus {\rm
e}^{-k\phi}
F_{m_1\cdots m_{k+1}} \Big] \,, \nn\\
g\, A_2 &=&  {\rm e}^{-[1+n/(d-2)]\phi} \Big[ {\cal Q}_m \oplus {\cal
P}_m
\oplus {\rm e}^{-k\phi} F_{m_1\cdots m_{k+1}} \Big] \,.
\eea
Hence, one identifies the same contributions to the $T$-tensor as in
\eqn{SS-potential}. There are some noteworthy features. One is that the
contributions of ${\cal Q}_m$, which contributes to both $A_1$ and
$A_2$,  cancel in the potential. The other is that the (negative)
contribution from $A_1$ to the potential is always compensated by the
(positive) contribution from $A_2$ in order that the final result
remains non-negative.
The contributions from the tensor fluxes can only be present for
multiple Scherk-Schwarz reductions where $n\geq k+1$. We conclude
that the Scherk-Schwarz $T$-tensors have a uniform behavior under
${\rm SO}(1,1)$, with a weight given by \eqn{so-weight}.  Furthermore
they transform as tensors under ${\rm SL}(n)$. We intend to exhibit
a number of examples of gaugings corresponding to hitherto new
Scherk-Schwarz reductions in \cite{dWSamtTrig2}. 

We already observed that the potential has only stationary points when
${\cal P}_m = F_{m_1\cdots m_{k+1}} =0$. Obviously, for large and
positive values of $\phi$ the potential tends to zero. In order to
have ${\cal P}_m=0$, there are restrictions on the generators $t_m$
involved in the Scherk-Schwarz reduction. Namely, at the stationary
point, the coset representative ${\cal V}_0$ and the generators $t_m$
should satisfy the condition that ${\cal V}^{-1}_0 t_m\,{\cal V}_0$
belongs to the Lie algebra associated with the isotropy group ${\rm
H}$. Since they thus belong to the same conjugacy class, we can
restrict the $t_m$ to a Cartan subalgebra associated with H, so that
the $t_m$ are anti-hermitean. From this observation we derive the
condition,
\be
{\cal V}_0 {\cal V}_0^\dagger \,t_m =  t_m\, {\cal V}_0 {\cal
V}_0^\dagger \,. 
\ee
Hence, we have mutually commuting generators $t_m$ which all commute
with ${\cal V}_0 {\cal V}_0^\dagger$. As the latter is a
positive-definite hermitean matrix, it can be diagonalized with
positive eigenvalues and the $t_m$ can simultaneously be
diagonalized. Furthermore, there is a gauge in which the coset
representative $\vv$ is hermitean, from which we deduce that the
stationary point is in fact part of a zero-potential valley, spanned
by the scalars that are invariant under the group generated by the
$t_m$. The fields that are not invariant under this group, must
therefore vanish at the stationary point. Of course, whether or not
the field strengths $F_{m_1\cdots m_{k+1}}$ vanish is a question that
is independent from the above considerations.

\section{The known gaugings in $d=4$ dimensions}
\setcounter{equation}{0}
In this section we demonstrate how the group-theoretical approach of
this paper allows us to straightforwardly establish the viability of
the known gaugings of maximal supergravity in 4 dimensions. The
strategy is to make an assumption about the gauge group ${\rm G}_g$,
or about the electric subgroup ${\rm G}_{\rm e}$ which contains ${\rm
G}_g$ as a subgroup, and then analyze the constraint
\eqn{constraint-912}. We do this by comparing the branching of the
${\bf 133}\times{\bf 56}$ representation under a given electric
subgroup ${\rm G}_{\rm e}$, to the representations that are allowed
for the $T$-tensor according to \eqn{constraint-912}. In this way we
can identify common representations in the product representation and
ultimately determine the possible assignment of the $T$-tensor to
representations of ${\rm G}_{\rm e}$. At the same time we deduce which
gauge fields and ${\rm E}_{7(7)}$ generators are involved in the
corresponding gauging. In a sequel we review the situation regarding
the known gaugings based on ${\rm G}_{\rm e}$ equal to ${\rm
SL}(8,\mathbb{R})$, ${\rm E}_{6(6)}\times {\rm SO}(1,1)$ and ${\rm
SU}^\star (8)$.

In the comparison and the identification of the various
representations, the projector $\mathbb{P}_{(912)}$, defined in
\eqn{E7-projectors}, plays a central role.  The part of the analysis
that involves its explicit form is done by means of the computer.
Because a direct handling of the $(56\cdot133)\times(56\cdot133)$
matrix $\mathbb{P}_{(912)}$ is rather cumbersome, we construct an
orthogonal basis of 912 embedding matrices spanning the ${\bf 912}$
representation. Putting some of the representations in the ${\bf 56}$
and the ${\bf 133}$ representations to zero, we analyze a system of
corresponding linear equations which then reveals the precise
branching rules.

We should stress that this approach is by no means limited to $d=4$
and can also be applied to gauged supergravity theories in other
dimensions. However, the analysis is most subtle for $d=4$ in view of
electric-magnetic duality and, in this section we restrict ourselves to
that case. To demonstrate our strategy, we will now proceed and
discuss the three classes of known gaugings.

\subsection{Gauge groups embedded in ${\rm SL}(8,\mathbb{R})$}
The first class of gaugings concerns gauge groups that can be embedded
into the ${\rm SL}(8,\mathbb{R})$ subgroup of \Exc7. The relevant
branching rules into ${\rm SL}(8,\mathbb{R})$ representations are,
\begin{eqnarray}
{\bf 56}&\rightarrow & {\bf 28}+\overline{\bf 28}\,, \nonumber\\
{\bf 133}&\rightarrow & {\bf 63}+{\bf 70}\,, \nonumber\\
{\bf 912}&\rightarrow & {\bf 36}+{\bf 420}+ \overline{\bf 36}+
\overline{\bf 420} \,,
\label{sl8decs}
\end{eqnarray}
where the ${\bf 28}$ representation in the first branching corresponds
to the gauge potentials and the conjugate $\overline{\bf 28}$
corresponds to the dual magnetic potentials, which cannot be included
in the gauging.  Recall that the embedding matrix $\Theta_M{}^\alpha$
living in the ${\bf 912}$ representation encodes the coupling of the
gauge fields to the \Exc7 generators (\ref{gauge-bein}),
(\ref{TTheta0}) and therefore its index $M$ refers to the
$\overline{\bf 28}$ representation that is conjugate to the
representation to which the gauge fields have been assigned. Note also
that, according to \eqn{TTheta}, the $T$-tensor transforms in the
conjugate representation (induced by the transformations of $\vv$), as
compared to the representation found for the embedding matrix
$\Theta_M{}^\alpha$. The branchings of products of the relevant
representations (\ref{sl8decs}) that belong to the ${\bf 912}$, and
thus identify acceptable representations of a $T$-tensor, is
conveniently summarized by the table below,
\begin{equation}
\begin{tabular}{|c||c|c|} \hline
&${\bf 28}$ &$\overline{\bf 28}$
\\
\hline
\hline
${\bf 63}$
&${\bf 36}+{\bf 420}$
&$\overline{\bf 36}+\overline{\bf 420}$
\\
\hline
${\bf 70}$
&$\overline{\bf 420}$
&${\bf 420}$
\\
\hline
\end{tabular}
\label{decblock}
\end{equation}
Because only one ${\bf420}$ representation appears in the branching of
the ${\bf912}$, the two ${\bf420}$ representations in \eqn{decblock}
must coincide, and so must the $\overline{\bf420}$
representations. This implies that, if the embedding matrix
$\Theta_M{}^\alpha$ had a contribution in the ${\bf420}$, it would
describe a coupling of the gauge fields to the generators in the~${\bf
70}$, but also, at the same time, induce a coupling of the dual gauge
fields to the generators in the ${\bf 63}$ representation of ${\rm
SL}(8,\mathbb{R})$. Since this is not possible in a local field theory
the embedding matrix must transform in the~$\overline{\bf 36}$
representation.  Indeed, according to (\ref{decblock}), the gauge
group generators are then contained in the adjoint ${\bf 63}$
representation of ${\rm SL}(8,\mathbb{R})$. Thus we conclude that all
possible gaugings for the Lagrangian in the ${\rm SL}(8,\mathbb{R})$
basis are defined by an embedding matrix in the $\overline{\bf 36}$
representation.  The $\overline{\bf 36}$ representation falls in
different conjugacy classes with respect to ${\rm SL}(8,\mathbb{R})$,
characterized by the eigenvalues of a symmetric $8\times 8$ matrix
$\theta_{AB}$, which can be taken equal to $\pm1$ or $0$. There are 44
nontrivial conjugacy classes, of which 24 correspond to inequivalent
gaugings. For $p$ eigenvalues $+1$, $q$ eigenvalues $-1$ and $r$
eigenvalues equal to 0, the resulting gauge group equals ${\rm
CSO}(p,q,r)$. The matrix $\theta_{AB}$ is invariant under this group.

We stress that the embedding matrix completely determines the
gauging. The embedding matrix $\Theta_{[AB]}{}^{\!C}{}_{\!D}
\propto \d^C_{[A}\,\theta^{~}_{B]D}$, where $\theta_{AB}$ denotes the
component of the embedding matrix in the $\overline{\bf 36}$
representation, defines gauge group generators in terms of the ${\rm
SL}(8,\mathbb{R})$ generators $t_A{}^{\!B}$ ({\it
cf.}~\eqn{gauge-gen})
\be 
t_{AB}= \Theta_{[AB]}{}^{\!C}{}_{\!D}  \,t_C{}^{\!D} 
\propto \theta^{~}_{D[A} \,t^{~}_{B]}{}^{\!D} \,.
\ee
Indeed these are the generators corresponding to ${\rm CSO}(p,q,r)$.
For instance, consider the case $p=8$, $q=r=0$. The embedding matrix
is then invariant under the ${\rm SO}(8)$ subgroup of ${\rm
SL}(8,\mathbb{R})$. There is just one ${\rm SO}(8)$ invariant
contraction, which corresponds to the ${\rm SO}(8)$ gauging of
\cite{deWitNic}. In the case that $p+q=8$ and $r=0$, the embedding
matrix is ${\rm SO}(p,q)$ invariant, and again, this identifies the
latter as the unique gauge group. In the case that $r\not=0$ we can
use a contraction of ${\rm SO}(p,q)$, which leaves the Lagrangian, and
in particular the $T$-tensor invariant. In this way we recover all the
${\rm CSO}(p,q,r)$ gaugings with $p+q+r=8$ found in
\cite{Hull} ({\it cf.} table~12 of~\cite{exhaustive}).
 
The ${\rm CSO}(p,q,r)$ gaugings with $r\not=0$ can also be obtained
from the ${\rm SO}(p+r,q)$ gauging by a contraction \cite{Hull}. In
order to show this let us briefly discuss the behaviour of the various
quantities under the diagonal subgroup, {\it i.e.} the ${\rm
SL}(8,\mathbb{R})$ matrices which are diagonal with eigenvalues
$\lambda_A$ (subject to the condition
$\Pi_{A=1}^8\,\lambda_A=1$). Under these transformations it follows
from \eqn{spelec} and \eqn{sumelec}, that (no summation over repeated
indices),
\bea
\label{diag-sl8}
A_\m^{AB} &\to& \lambda_A \lambda_B \,A_\m^{AB}\,, \nonumber \\
u_{ij}{}^{\!AB} \pm v_{ijAB} &\to& (\lambda_A \lambda_B)^{\pm 1} 
\,(u_{ij}{}^{\!AB} \pm v_{ijAB}) \,.
\eea
It is easy show that the Lagrangian \eqn{L3} is indeed invariant under
these transformations, but the terms induced by the gauging ({\it
i.e.} both in the covariant derivative and in the $T$-tensor) are not.
Decompose the 28 ${\rm SO}(p+r,q)$ generators as
\be 
\label{so(p+r,q)}
\pmatrix{{\sf A}_{p+q} & {\sf C} \eta \cr\noalign{\vskip2mm}
         -\eta {\sf C}^{\rm T} & {\sf B}_r \cr} 
\;,
\ee 
where ${\sf A}_{p+q}^{\rm T}= -\eta {\sf A}\eta$ and ${\sf B}^{\rm T}
= -{\sf B}$ with $\eta$ the ${\rm SO}(p,q)$ invariant metric. 
Applying the transformation \eqn{diag-sl8} has the effect of
both rescaling the embedding matrix and the generators.

Consider a special transformation with $\lambda_A=\lambda$ for the
first $p+q$ eigenvalues and $\lambda_A=\sigma$ for the last $r$
eigenvalues, so that $\lambda^{p+q}\,\sigma^r=1$. In that case ${\sf
A}$ and ${\sf B}$ remain unchanged whereas one of the the off-diagonal
blocks is multiplied by $\sigma/\lambda$ and the other one by its
inverse. The contraction of these generators with the gauge fields
$A_\m^{AB}$, or the quantities $(u_{ij}{}^{\!AB} + v_{ijAB})$ as in
the covariant derivative and the $T$-tensor, respectively, introduces
additional factors $\lambda^2$, $\sigma\lambda$ or $\sigma^2$,
depending on the index values. Performing now the singular limit
$\sigma\to 0$ on finds that the contraction is proportional to
$\lambda^2$ (which can be absorbed in the coupling constant, so that
the covariant derivative and the $T$-tensor remain finite) times the
generators \eqn{so(p+r,q)} with ${\sf B}$ and one of the off-diagonal
blocks suppressed. The dimension of the contracted gauge group ${\rm
CSO}(p,q,r)$ is thus equal to $\ft12 (p+q)(p+q-1+2r)$, and the
corresponding algebra consists of the generators ${\sf A}$ and
$r(p+q)$ nilpotent generators residing in the off-diagonal block ${\sf
C}$.

\subsection{Gauge groups embedded in the ${\rm E}_{6(6)}$ basis}
The second class of gaugings concerns gauge groups that can be found
in the ${\rm E}_{6(6)}$ basis defined in section \ref{coset}. Again we
start by giving the relevant branchings of the various ${\rm
E}_{7(7)}$ representations with respect to ${\rm E}_{6(6)}\times {\rm
O}(1,1)$. The branching of the ${\bf 56}$, ${\bf 133}$ and ${\bf 912}$
representations are given below (the subscript refers to the ${\rm
SO}(1,1)$ weight),
\begin{eqnarray}
{\bf 56}&=&   \overline{\bf 27}_{-1} +{\bf
1}_{-3}+    {\bf 27}_{+1}+{\bf 1}_{+3} \,,\nn\\ 
{\bf 133}&=& {\bf 78}_0 + \overline{\bf 27}_{+2} + {\bf 27}_{-2}+ {\bf
1}_0\,,
\nn\\
{\bf 912}&=& \overline{{\bf 351}}_{-1}+ {\bf 351}_{+1}+
\overline{\bf 27}_{-1}+ {\bf 27}_{+1}+ {\bf 78}_{-3}+{\bf 78}_{+3} \,,
\label{Tdec45}
\end{eqnarray}
where in the second line ${\bf 78}_0$ and ${\bf 1}_0$ represent the
adjoint representations of ${\rm E}_{6(6)}$ and of ${\rm O}(1,1)$,
respectively. Just as before, we can conveniently summarize the
branchings of $({\bf 56}\times {\bf 133})\cap{\bf 912}$ in a table,
\begin{equation}
\begin{tabular}{|c||c|c|c|c|} \hline
&$\overline{\bf 27}_{-1}$&${\bf 1}_{-3}$  & ${\bf 27}_{+1}$
&${\bf 1}_{+3}$
\\
\hline\hline
${\bf 78}_0$
&$\overline{\bf 351}_{-1}+ \overline{\bf 27} _{-1}$
&${\bf 78} _{-3}$
&$ {{\bf 351}}_{+1}+ {\bf {27}} _{+1}$
&${\bf 78} _{+3}$
\\
\hline
${\bf 27}_{-2}$
&${\bf 78} _{-3}$
&
&$\overline{\bf 351}_{-1}+ \overline{\bf 27} _{-1}$
&${\bf 27}_{+1}$
\\
\hline
$\overline{\bf 27}_{+2}$
&$ {{\bf 351}}_{+1}+ {\bf{27}} _{+1}$
& $\overline{\bf 27} _{-1}$
&${\bf 78} _{+3}$
&
\\
\hline
${\bf 1}_0$
& $\overline{\bf 27} _{-1}$
&
& ${\bf 27}_{+1}$
&
\\ \hline
\end{tabular}
\label{decblock-e6}
\end{equation}
Again equivalent representations in this table must coincide, since
the ${\bf912}$ contains just a single copy of each. With a similar
reasoning as above, it follows that viable gaugings involve the gauge
fields (in the $\overline{\bf 27} _{-1}+{\bf 1} _{-3}$ representation)
coupling to \Exc7 generators belonging to the ${\bf 78} _{0}+{\bf
\overline{27}}_{+2}$ representation and the corresponding embedding
matrix is contained the ${\bf 78}_{+3}$ representation (we recall that
the embedding matrix is assigned to the representation that is
conjugate with respect to one to which the gauge fields have been
assigned).

This completely determines all possible gaugings in this basis. The
gauge field in the ${\bf 1}_{-3}$ representation couples to an element
of the adjoint representation of \Exc6, whereas (part of) the gauge
fields in the $\overline{\bf 27}_{-1}$ representation couple to
generators in the $\overline{\bf 27}_{+2}$ representation of ${\rm
E}_{6(6)}\times{\rm SO}(1,1)$.  This gauging is not new and has an
interpretation as a Scherk-Schwarz reduction from $d=5$ maximal
supergravity. Indeed, the ${\rm SO}(1,1)$ weights of the gauge fields
are consistent with \eqn{torus-scale}): the graviphoton transforms in
the ${\bf 1}_{-3}$ representation and the 27 gauge fields from the
five-dimensional theory transform in the $\overline{\bf 27}_{-1}$
representation. In the Scherk-Schwarz reduction the graviphoton
couples to one of the \Exc6 generators, whereas the remaining 27 gauge
fields couple to generators in the $\overline{\bf 27}_{+2}$
representation. This interpretation is also confirmed by the fact that
the $T$-tensor transforms in the ${\bf 78}_{-3}$ representation
(remember that the representation is conjugate to the one for the
embedding matrix), in accord with \eqn{so-weight}. The form of the
gauge group generators follows from \eqn{e66basis} and the dimension
of the gauge group follows from the rank of the \Exc6 generator that
has been selected. Hence we reproduce the
result~\cite{AndDauFerrLle}. This particular Scherk-Schwarz reduction
has been worked out in~\cite{sezginvN}.

These above gaugings are thus defined by a family of embedding
matrices which associate the graviphoton generator $X_0$ with one of
the $78$ generators $t_0$ in \Exc6, and the remaining gauge generators
$X_\Lambda$ with the nilpotent generators denoted by $t_\Lambda$ in
the $\overline{\bf 27} _{+2}$ representation. {From} $[t_0,t_\Lambda]
= M_\Lambda{}^\Sigma\, t_\Sigma$, it follows that the embedding matrix
can then be represented as follows,
\begin{eqnarray} 
X_0&=&t_0\,,\qquad  X_\Lambda = M_\Lambda{}^\Sigma t_\Sigma\,,
\label{embefla}
\end{eqnarray}
so that the Lie algebra based on the generators $\{X_0,\,X_\Lambda\}$
takes the form,
\begin{eqnarray} 
[X_0,\,X_\Lambda]&=& M_\Lambda{}^\Sigma
X_\Sigma\,, \qquad [X_\Lambda,\,X_\Sigma]=0 \,.
\end{eqnarray}
The null vectors of the matrix $M$ correspond to gauge fields that do
not participate in the gauging. As we discussed in section~5, when
$t_0$ is a generator belonging to the compact subgroup of \Exc6, the
corresponding potentials have stationary points with Minkowski ground
states. This subgroup is equal to the group ${\rm USp}(8)$, which has
conjugacy classes described in terms of four real parameters
$m_1,\ldots,m_4$ ({\it cf.} \cite{sezginvN}). Generically the matrix
$M$ has then three zero eigenvalues and 24 eigenvalues equal to the
linear combinations $\pm m_i\pm m_j$ with $i>j$ taking the values
$1,\ldots,4$. These are the weights of the ${\bf 27}$ representation
of ${\rm USp}(8)$ which determine the mass matrix of the vector
fields.  The gauge group dimension takes odd values between 13 and
25. The 13-dimensional gauge group is equal to ${\rm CSO}(2,0,6)$.

\subsection{Gauge groups embedded into ${\rm SU}^\star (8)$}
As we discussed in section~2, ungauged Lagrangians exist with ${\rm
SU}^\star (8)$ symmetry \cite{Hull2}. The group ${\rm SU}^\star (2n)$
is defined as the group of complex $2n\times 2n$ matrices ${\sf U}$
satisfying:
\begin{eqnarray}
{\sf U}{\sf J}&=&{\sf J}{\sf U}^\star \;,
\end{eqnarray}
where ${\sf J}$ is the ${\rm Sp}(2n)$ invariant form. The compact
subgroup is equal to ${\rm USp}(2n)$, and the real subgroup is equal
to ${\rm SO}^\star (2n)$. Finally the groups ${\rm CSO}^\star
(2p,2n-2p)$ can be obtained by a contraction of ${\rm SO}^\star (2n)$.

In this basis ${\rm SU}^\star (8)$ acts block-diagonally on the gauge
fields and their dual potentials. The relevant ${\rm E}_{7(7)}$
representations decompose with respect to ${\rm SU}^\star (8)$
precisely as in (\ref{sl8decs}).  However, in contradistinction to
${\rm SL}(8,\mathbb{R})$, the group ${\rm SU}^\star (8)$ corresponds
to a real form of ${\rm SL}(8,\mathbb{C})$ with $36$ generators
defining the ${\rm USp}(8)$ subgroup and $27$ generating the
noncompact components of the group. The maximal compact subgroup can
be defined by ${\rm USp}(8)={\rm SU}^\star (8)\cap {\rm
E}_{6(6)}$. The embedding of ${\rm SU}^\star(8)$ inside ${\rm
E}_{7(7)}$ can be conveniently described by decomposing the adjoint of
the latter with respect to~${\rm USp}(8)$:
\begin{eqnarray}
 {\bf 133} &\rightarrow &  {\bf 36} + {\bf
42} + {\bf 1} + {\bf 27} _{\rm nc}+ {\bf
27} _{\rm c}\,, \nonumber\\
{\rm Adj}({\rm SU}^\star (8))&\rightarrow& {\bf 36} + {\bf
27} _{\rm nc}\,, 
\end{eqnarray}
where ${\bf 27}_{\rm nc}$ denotes the noncompact linear combinations
of the nilpotent generators ${\bf 27}_{-2}$ and ${\overline{\bf
27}}_{+2}$ from (\ref{Tdec45}); ${\bf 27}_{\rm c}$ denotes the
corresponding compact combination. The ${\bf 70}$ in the decomposition
of the ${\rm Adj}({\rm E}_{7(7)})$ representation with respect to
${\rm SU}^\star (8)$ consists therefore of the ${\rm USp}(8)$
representation $ {\bf 42} + {\bf 1} + {\bf 27} _{\rm c}$. The
transformation from the ${\rm E}_{6(6)}$ basis to the ${\rm SU}^\star
(8)$ basis was defined in (\ref{e6sust}).

The decomposition of the $ {\bf 912} $ with respect to ${\rm SU}^\star
(8)$ takes the same form as for the ${\rm SL}(8,\mathbb{R})$
case. Therefore it follows that the embedding matrix must belong to
the ${\bf 36}$ representation of ${\rm SU}^\star (8)$. This
representation has four nontrivial conjugacy classes, depending on the
rank of the embedding matrix. For a nonsingular embedding matrix the
gauge group equals the real subgroup ${\rm SO}^\star (8)$. Other
gaugings correspond to the groups ${\rm CSO}^\star (2p,8-2p)$ and can
be described by an appropriate contraction, just as as described
earlier for the ${\rm SL}(8,\mathbb{R})$ basis and analyzed
in~\cite{Hull2}.

\section{Some gaugings in $d=5$ dimensions}
As yet another application, we analyze some of the gaugings in $d=5$
maximal supergravity. In this case there are no subtleties related to
electric-magnetic dualities and the search for viable gauge groups
should be based on arbitrary subgroups of \Exc6 without the need for
referring to a specific basis. We consider several
classes. First we assume that the gauge group is a subgroup of the 
${\rm SL}(2,\mathbb{R})\times {\rm SL}(6,\mathbb{R})$ maximal subgroup
of \Exc6. Then we consider the case where the gauge group is embedded
in the non-semisimple extension ${\rm SO}(5,5)\times {\rm SO}(1,1)$,
which is also a maximal subgroup of~${\rm E}_{6(6)}$. 

As proven in section~4 the embedding matrix for $d=5$ dimensions must
belong to the ${\bf 351}$ representation of \Exc6, {\it i.e.}, 
\begin{eqnarray}
\label{constraint-351}
\mathbb{P}_{(351)\Lambda}{}^{\!a\, \Sigma}{}_{\!b}\,\Theta_\Sigma{}^b
&=& \Theta_\Lambda{}^a\,,
\end{eqnarray}
where the projection operator was defined in \eqn{E6-projectors}. With
respect to the ${\rm SL}(2,\mathbb{R})\times {\rm SL}(6,\mathbb{R})$
of \Exc6, the vector gauge fields, the \Exc6
generators and the embedding matrix decompose according to, 
\begin{eqnarray}
\overline{\bf 27} &\rightarrow& ({\bf 1},\overline{{\bf 15}}) +
({\bf 2},{\bf 6}) \,,
\nonumber\\
{\bf 78} &\rightarrow&
({\bf 1},{\bf 35}) + ({\bf 3},{\bf 1})
+ ({\bf 2},{\bf 20})\,,
\nonumber\\
{\bf 351} &\rightarrow&
({\bf 1},\overline{\bf 21}) + ({\bf 3},{\bf 15}) 
+ ({\bf 2},\overline{\bf 84})
+ ({\bf 2},\overline{\bf 6}) + ({\bf 1},\overline{\bf 105})\,,
\label{decblock-sl6}
\end{eqnarray}
respectively.  The table below summarizes how the embedding matrix
couples the vector fields to the generators (we recall that the
embedding matrix is assigned to the $({\bf 27}\times{\bf 78})\cap{\bf
351}$ representation),
\begin{equation}
\begin{tabular}{|c||c|c|} \hline
&$({\bf 1},{\bf 15})$
&$({\bf 2},\overline{\bf 6})$
\\ \hline\hline
$({\bf 1},{\bf 35})$
&$({\bf 1},\overline{\bf 21}) + ({\bf 1},\overline{\bf 105}) $
&$({\bf 2},\overline{\bf 6})+({\bf 2},\overline{\bf 84}) $
\\ \hline
$({\bf 3},{\bf 1})$
&$({\bf 3},{\bf 15})$
&$({\bf 2},\overline{\bf 6})$
\\ \hline
$({\bf 2},{\bf 20})$
&$({\bf 2},\overline{\bf 6})+({\bf 2},\overline{\bf 84}) $
&$({\bf 3},{\bf 15}) + ({\bf 1},\overline{\bf 105}) $
\\ \hline
\end{tabular}
\end{equation}
where again equivalent representations in the table must be
identified. Only the first two rows are, however, acceptable because
those belong to the generators of ${\rm SL}(2,\mathbb{R})\times {\rm
SL}(6,\mathbb{R})$ and we assumed that the gauge group was embedded in
this group. This leaves only one possible representation assignment
for the embedding matrix, namely it should belong to the $({\bf
1},\overline{\bf 21})$ representation and only the vector fields
transforming in the $({\bf 1},\overline{\bf 15})$ representation are
involved in the gauging and couple to the generators in the adjoint
representation of ${\rm SL}(6,\mathbb{R})$. The charged vector fields
in the $({\bf 2},{\bf 6})$ cannot participate in the gauging and must
be dualized into antisymmetric tensor fields. The gaugings are again
completely determined and correspond to the conjugacy classes of the
$({\bf 1},\overline{\bf 21})$ representation. They lead to the ${\rm
CSO}(p,q,r)$ gauge groups with $p+q+r=6$ found by
\cite{GunaRomansWarner,AndCordFreGual}. Apart from the conversion into
tensor fields, this is entirely analogous to
the discussion in $d=4$ dimensions for the ${\rm SL}(8,\mathbb{R})$
basis. 

A second application is based on ${\rm SO}(5,5)\times {\rm
SO}(1,1)$. This semisimple group is not a maximal subgroup of \Exc6,
but it becomes maximal upon including 16 additional nilpotent
generators transforming in the $\overline {\bf 16}_{+3}$
representation. We assume that the gauge group will be a subgroup of
this non-semisimple maximal subgroup.  The decompositions of the
relevant ${\rm E}_{6(6)}$ representations with respect to the ${\rm
SO}(5,5)\times {\rm SO}(1,1)$ subgroup is given by,
\begin{eqnarray}
\overline{\bf 27} &=& \overline{\bf 16}_{-1}+{\bf 10}_{+2}+{\bf
1}_{-4}\,,\nonumber\\ 
{\bf 78} &=& {\bf 45}_{0}+{\bf 1}_{0}+ {\bf 16}_{-3}+\overline{{\bf
16}}_{+3}\,,\nonumber\\
{\bf 351} &=& {\bf 144}_{+1}+
{\bf 16}_{+1}+{\bf 45}_{+4}+{\bf 120}_{-2}+{\bf 10}_{-2}+{\bf
\overline{16}}_{-5}\,.
\end{eqnarray}
The couplings induced
by an embedding matrix solving (\ref{constraint-351}) are shown below, 
\begin{equation}
\begin{tabular}{|c||c|c|c|} \hline
&${\bf 16}_{+1}$
&${\bf 10}_{-2}$
&${\bf 1}_{+4}$
\\
\hline\hline
${\bf 45}_{0}$
& ${\bf 144}_{+1}+ {\bf 16}_{+1}$
&${\bf 10}_{-2}+{\bf 120}_{-2}$
& ${\bf 45}_{+4}$
\\
\hline
${\bf 1}_{0}$
&${\bf 16}_{+1}$
&${\bf 10}_{-2}$
&
\\
\hline
${\bf 16}_{-3}$
&${\bf 120}_{-2}+ {\bf 10}_{-2}$
&${\overline{\bf 16}}_{-5} $
&${\bf 16}_{+1}$
\\
\hline
$\overline{{\bf16}}_{+3}$
&${\bf 45}_{+4}$
&${\bf 144}_{+1}+ {\bf 16}_{+1}$
&\\
\hline
\end{tabular}
\end{equation}
Again equivalent representations for the embedding matrix should be
identified as they appear with multiplicity one in the ${\bf 351}$
representation. Furthermore, the generators belonging to the ${\bf
16}_{-3}$ cannot be involved in the gauging, as they do not belong to
the maximal subgroup that we have chosen. Therefore only two
representations are allowed for the embedding matrix, namely the ${\bf
144}_{+1}$ and the ${\bf 45}_{+4}$ representation. As we will outline
below, two particular classes corresponding to each of these
representations can be immediately identified. No gaugings have been
worked out so far with an embedding matrix that contains components
from both representations. We will return to this elsewhere.

When the embedding matrix belongs to the ${\bf 144}_{+1}$
representation, one can consider gauged supergravity in $d=6$
dimensions, whose embedding matrix must be in the ${\bf 144}$
representation of the ${\rm SO}(5,5)$ duality group ({\it cf.}
\eqn{embedd-d}). Upon dimensional
reduction on $S^1$, one finds a $T$-tensor in the ${\bf 144}_{-1}$
representation, which is indeed conjugate to the representation of the
embedding matrix. However, not too much is known about gaugings for
the $d=6$ theory (see, for example, \cite{Cowdall})

When the embedding matrix belongs to the ${\bf 45}_{+4}$
representation we are dealing with a Scherk-Schwarz reduction from
$d=6$ dimensions, where ungauged maximal supergravity is invariant
under ${\rm SO}(5,5)$ duality. To verify this, first consider the
representations for the vector fields whose six-dimensional origin is
as follows.  The ${\bf 1}_{-4}$ vector field corresponds to the
graviphoton, the $\overline{\bf 16}_{-1}$ vector fields originate from
the 16 $d=6$ vector fields, and the ${\bf 10}_{+2}$ vector fields
originate from the 10 $d=6$ tensor fields. The ${\rm SO}(1,1)$ weights
are in accord with the results given in
\eqn{torus-scale}. Regarding the tensors, we note that a tensor
$A_{MN}$ in 6 dimensions leads to a vector $A_{\m 6}$ and a tensor
$A_{\m\n}$ in 5 dimensions, which, according to
\eqn{torus-scale} have weights $+2$ and $-2$, respectively. However,
a 2-rank tensor gauge field in 5 dimensions can be dualized into a
vector field with opposite weight, so that all the vectors originating
from tensor gauge fields in 6 dimensions carry the same ${\rm
SO}(1,1)$ weight equal to $+2$. The $T$-tensor in this reduction must
be in the ${\bf 45}_{-4}$ representation, which is indeed conjugate to
the representation of the embedding matrix identified above.

Just as in the four-dimensional case discussed in section~6.2, one
identifies the representation ${\bf 45}_{+4}$ with the complete family
of five-dimensional gauge groups generated by
$\{X_0,\,X_p\}_{p=1,\dots,16}$, characterized by associating the
generator $X_0$ which couples to the graviphoton in the ${\bf 1}_{-4}$
representation with a generator $t_0$ of ${\rm SO}(5,5)$. The
remaining generators $X_p$ couple to the gauge fields in the
$\overline{\bf 16}_{-1}$ representation and are associated with the
nilpotent generators $t_p$ in the $ {\bf \overline{16}}_{+3}$
representation. {From} $[t_0,t_p]= M_p{}^q \,t_q$ one finds the
embedding matrix given by
\begin{eqnarray}
X_0&=&t_0\,, \qquad  X_p = {M}_p{}^q \, t_q\,,
\label{embefla5}
\end{eqnarray}
with the following gauge algebra generated by $\{X_0,\,X_p\}$, 
\begin{eqnarray}
[X_0,\,X_p]&=& {M}_p{}^q X_q\,, \qquad [X_p,\,X_q]=0\,.
\end{eqnarray}
{From} (\ref{embefla5}) it follows that the null vectors of $M_p{}^q$
correspond to gauge fields that do not participate in the gauging and
remain abelian. Obviously the maximal dimension of the gauge group is
equal to $17$. This means that at least 10 of the $27$ vector fields
of the $d=5$ theory must be converted to charged tensor fields; at any
rate these include the vectors in the ${\bf 10}_{+2}$.

\begin{table}
\begin{center}
\begin{tabular}{l |  c c c r r}\hline
field&
spin/helicity & 
{\rm H}-rep & masses & $\#$ & dof's  \\
\hline
graviton &
$5$ & ${\bf (1,1)} $ & 0 & 
\begin{tabular}{c}1\end{tabular} &\begin{tabular}{c}5\end{tabular} 
\\[1mm]
gravitini &
$(3,2)$ & ${\bf (4,1)} $ & $|m_r|$ & 
\begin{tabular}{c}2\end{tabular} &\begin{tabular}{c}24\end{tabular} 
\\
 & $(2,3)$  & ${\bf (1,4)} $ & $|\tilde{m}_r|$& 
\begin{tabular}{c}2\end{tabular} &\begin{tabular}{c}24\end{tabular} 
\\[1mm]
vectors &
$(2,2)$   & ${\bf (4,4)}$ & $|m_r\pm \tilde{m}_s|$ & 
\begin{tabular}{c}2\end{tabular} &\begin{tabular}{c}64\end{tabular} 
\\[1mm]
    & 3  & ${\bf (1,1)}$ & $0$ & 
\begin{tabular}{c}1\end{tabular} &\begin{tabular}{c}3\end{tabular} 
\\[3mm]
tensors &
  $(3,1)$ & ${\bf (5,1)}$ & $\cases{|m_1\pm m_2|\cr 0}$& 
\begin{tabular}{c}2\\1\end{tabular} 
& \begin{tabular}{c}12\\3\end{tabular} \\ [3mm]
& $(1,3)$   & ${\bf (1,5)}$ & $\cases{|\tilde{m}_1\pm \tilde{m}_2|\cr 0}$&
\begin{tabular}{c}2\\1\end{tabular}   
&  \begin{tabular}{c}12\\3\end{tabular} \\ [6mm]
spinors &
$(1,2)$ & ${\bf (4,5)} $ & $\cases{|m_r\pm
\tilde{m}_1\pm\tilde{m}_2|\cr|m_r|}$& 
\begin{tabular}{c}2\\2\end{tabular}  & 
\begin{tabular}{c}32\\8\end{tabular} \\[3mm]
&$(2,1)$  &${\bf (5,4)} $ & $\cases{|m_1\pm m_2\pm
\tilde{m}_s|\cr|\tilde{m}_s|}$& 
\begin{tabular}{c}2\\2\end{tabular}  & 
\begin{tabular}{c}32\\8\end{tabular} \\[5mm] 
scalars &
$0$   & ${\bf (1,1)}$ & $0$ &  
\begin{tabular}{c}1\end{tabular} &\begin{tabular}{c}1\end{tabular} 
\\[1mm]
&   & ${\bf (5,5)}$ & $\cases{|m_1\pm
m_2\pm\tilde{m}_1\pm\tilde{m}_2|\cr|m_1\pm m_2|\cr
|\tilde{m}_1\pm \tilde{m}_2|\cr 0}$ & 
\begin{tabular}{c}2\\2\\2\\1\end{tabular} &
\begin{tabular}{c}16\\4\\4\\1\end{tabular}\\
{~}\\
\hline 
\end{tabular}
\end{center}
\caption{\small Mass spectrum of maximal $d=5$ gauged supergravity
with an embedding matrix in the $({\bf 10},{\bf 1})+({\bf 1},{\bf
10})$ representation of ${\rm H}= {\rm USp}(4)\times {\rm
USp}(4)$.  Massless states transform under ${\rm SU}(2)$
helicity rotations; the dimension of the corresponding representation
is indicated by a single number. Massive states transform under
${\rm SU}(2)\times {\rm SU}(2)$ spatial rotations, so that their
spin-content is characterized by two numbers.}
\label{mtab}
\end{table}

When $X_0$ belongs to the Lie algebra associated with the maximal
compact subgroup of ${\rm SO}(5,5)$, which is equal to the rank-4
group ${\rm H}= {\rm USp}(4)\times {\rm USp}(4)$, it can be described
by four real parameters $m_1, m_2$ and $\tilde m_1,\tilde m_2$, which
denote the eigenvalues of $X_0$ in the ${\bf (4,1)}$ and in the ${\bf
(1,4)}$ representation, respectively. As discussed in
section~\ref{Scherk-Schwarz} the corresponding potential (which is
nonnegative) has minima with zero cosmological constant. For generic
values of the four parameters we can determine the physical masses,
collected in table~\ref{mtab}.

With respect to ${\rm H}={\rm USp}(4)\times {\rm USp}(4)$ the
gravitini transform in the ${\bf (4,1)}+ {\bf (1,4)}$ and the
spinors in the ${\bf (4,5)}+ {\bf (5,4)}+{\bf (4,1)}+
{\bf (1,4)}$ representations, respectively. The spinors
corresponding to the ${\bf (4,1)}+ {\bf (1,4)}$ representation are
absorbed into the massive gravitini through a
super-Brout-Englert-Higgs mechanism. The masses of the gravitini are
given by the eigenvalues of the matrix describing the action of $X_0$
on this representation.  The 16 vector fields and the 10 tensor fields
transform under ${\rm H}$ in the ${\bf (4,4)}$ and in the ${\bf
(5,1)}+{\bf (1,5)}$ representations, respectively. We observe that
there are two neutral, massless tensor fields which can be
consistently dualized back into vector fields. The scalar fields
transform under {\rm H} in the ${\bf (1,1)}+{\bf (4,4)}+{\bf (5,5)}$
representation. The scalars in the ${\bf (4,4)}$ arise from the
internal components of the vector fields in six dimensions. They are
absorbed into the massive gauge vectors.

We would like to conclude this discussion with a remark on the
four-dimensional models obtained by simple dimensional reduction on a
circle from five-dimensional gauged maximal supergravity. As pointed
out in section~6.2, dimensional reduction of the ungauged
five-dimensional theory leads to the 28 four-dimensional vector fields
described in~(\ref{sl8e6}). On the other hand, the ${\rm CSO}(p,q,r)$
gaugings in five dimensions break the ${\rm E}_{6(6)}$ invariance, not
only as a consequence of the minimal couplings, but also because the
vector fields in the ${\bf (2,6)}$ have to be dualized into tensor
fields in the ${\bf (2,\overline{6})}$. As a consequence of this
dualization, dimensional reduction of the ${\rm CSO}(p,q,r)$ gauged
theories gives rise to four-dimensional vector fields transforming as
in~(\ref{sl6sl2}), {\it i.e.}, it leads to four-dimensional gaugings
in the ${\rm SL}(2,\mathbb{R})\times {\rm SO}(1,1)\times {\rm
SL}(6,\mathbb{R})$ basis. Since the graviphoton is ungauged, it may
still be dualized: $({\bf 1},{\bf 1})_{-3}\rightarrow ({\bf 1},{\bf
1})_{+3}$. This yields a gauged model in the ${\rm SL}(8,\mathbb{R})$
basis, {\it cf.}~(\ref{28-sl8}), with gauge group ${\rm CSO}(p,q,r+2)$
that has been considered in section~6.1. This has also been discussed
in~\cite{Hull2}.

\vspace{8mm}

\noindent
{\bf Acknowledgement}\\
\noindent
We thank H.~Nicolai and S.~Ferrara for useful discussions. This work
is partly supported by EU contracts HPRN-CT-2000-00122 and
HPRN-CT-2000-00131. M.T.\ is supported by a European Community Marie
Curie Fellowship under contract HPMF-CT-2001-01276.

\bigskip

\appendix
\section{Projectors on the embedding matrix: a general discussion}
\label{app-A}
\setcounter{equation}{0}
In this appendix, we explicitly construct the projectors onto the
irreducible representations in the tensor product of the fundamental
with the adjoint representation of an arbitrary simple group G. These
projectors for ${\rm G}=$\Exc7, and ${\rm G}=$\Exc6 have been used in
the main text to correctly identify the embedding matrices in $d=4$
and $d=5$, respectively.

Let us assume that the product of a fundamental representation ${\bf
D(\Lambda)}$ times the adjoint decomposes in the direct sum of ${\bf
D(\Lambda)}$ plus two other representations, ${\bf D_1}$ and~${\bf
D_2}$, 
\begin{eqnarray}
{\bf D(\Lambda)}\times {\bf Adj(G)}\rightarrow
{\bf D(\Lambda)}+{\bf D_1}+{\bf D_2}  \;.
\label{gendec}
\end{eqnarray}
As far as the lowest-dimensional fundamental representation in
concerned, the above branching rule holds true for any simple group
with the exception of ${\rm E}_8$ for which the fundamental coincides
with the adjoint representation.  The branching also holds for
orthogonal groups when the fundamental representation is replaced by
the spinor representation .  Denote $d_\Lambda={\rm dim}({\bf
D(\Lambda)})$, $d={\rm dim}({\rm G})$, and $\{t^\alpha\}$
($\alpha=1,\dots , d$) the generators of G in the ${\bf D(\Lambda)}$
representation. Furthermore, let $C_\theta,\,C_\Lambda$ be the
Casimirs of the adjoint and fundamental representations,
respectively. We define the invariant matrix $\eta^{\alpha\beta}={\rm
Tr}(t^\alpha t^\beta)$ and use it to rise and lower the adjoint
indices; it is related to the Cartan-Killing metric
$\kappa^{\alpha\beta}$ by
\begin{eqnarray}
\kappa^{\alpha\beta}&=&
\frac{d}{C_\Lambda d_\Lambda}\, \eta^{\alpha\beta}\;.
\end{eqnarray}
Using the definition of the Casimir operator, 
$C_\Lambda\,\bfone_{d_\Lambda} =\kappa_{\alpha\beta} t^\alpha
t^\beta$, we have the following relation
\begin{eqnarray}
f_{\alpha\beta}{}^\gamma\, f^{\alpha\beta}{}_\sigma
&=&-\frac{d}{d_\Lambda}\, C_{\rm r}\,\delta_{\sigma}^\gamma
\;,\qquad
\mbox{with} \quad
C_{\rm r}=\frac{C_\theta}{C_\Lambda}=
\frac{d_\Lambda}{d}\frac{g^{\vee}}{\tilde{I}_\Lambda}
\;,
\end{eqnarray}
where $g^{\vee}$ is the dual Coxeter number and $\tilde{I}_\Lambda$
is the Dynkin index of the fundamental representation.  In the simply
laced case there is a useful formula:
\begin{eqnarray}
C_{\rm r}&=&\frac{d_\Lambda}{d}\left(
\frac{d}{r}-1\right)\frac{1}{\tilde{I}_\Lambda}
\;,
\end{eqnarray}
with $r$ the rank of G.

Denote the projectors on the representations in (\ref{gendec}) by
$\mathbb{P}_{{\bf D(\Lambda)}},\,\mathbb{P}_{{\bf D_1} },\,
\mathbb{P}_{{\bf D_2} }$ which sum to the identity on ${\bf
D(\Lambda)}\times {\bf Adj(G)}$.  These three projectors can be
expressed in terms of three independent objects, namely:
\begin{eqnarray}
\mathbb{P}_{{\bf D(\Lambda)M}}{}^{\alpha N} {}_\beta
&=&\frac{d_\Lambda}{d}\, (t^\alpha t_\beta)_M{}^N\,,\nonumber\\
\mathbb{P}_{{\bf D_1}M }{}^{\alpha N} {}_\beta &=& a_1 \,\delta^\alpha
{}_{\!\beta}\,\delta_M{}^N+a_2 \, (t_\beta t^\alpha)_M{}^N+a_3\,
(t^\alpha 
t_\beta)_M{}^N\,,\nonumber\\
\mathbb{P}_{{\bf D_2}M }{}^{\alpha N} {}_\beta&=&
(1-a_1)\,\delta^\alpha{}_{\!\beta}\, \delta_M{}^N-a_2\, (t_\beta
t^\alpha)_M{}^N-({d_\Lambda/d}+a_3)\,(t^\alpha t_\beta)_M{}^N
\;,
\label{projectors}
\end{eqnarray}
with constants $a_1, a_2, a_3$. Making use of the fact that only three
representations appear in the decomposition (\ref{gendec}), these
coefficients may be determined by computing the contractions of
various products of the projectors (\ref{projectors}). This yield
\begin{eqnarray}
a_1 &=&\frac{d_\Lambda\left(4+(C_{\rm r}-4)
d)\right)+\Delta\left((C_{\rm r}-2)
d-2)\right)}{\left(10+d(C_{\rm r}-8)+d^2 (C_{\rm r}-2)\right)
d_\Lambda}\,, \nonumber\\
a_2 &=&-\frac{2\left(4+(C_{\rm r}-4)
d)\right)\left((d-1)d_\Lambda-2\Delta\right)}{\left(10+d(C_{\rm
r}-8)+d^2 (C_{\rm r}-2)\right) C_{\rm r} d }\,, \nonumber\\
a_3 &=&\frac{-d_\Lambda\left(4+(C_{\rm r}-4)
d)\right)\left(2+(C_{\rm r}-2)d\right)+\Delta\left(16(d-1)-10 (d-1)
C_{\rm r}+C_{\rm r}^2 d \right)}{\left(10+d(C_{\rm r}-8)+d^2 (C_{\rm
r}-2)\right) C_{\rm r} d } \;,
\nonumber
\end{eqnarray}
with $\Delta=\dim({\bf D_1})$. Moreover, $\Delta$ is determined to be
\begin{eqnarray}
\Delta &=& \frac{d_\Lambda}{2}\left[d-1+\frac{\sqrt{C_{\rm
r}}\left(10+d(C_{\rm r}-8)+d^2 (C_{\rm r}-2)\right) }{\sqrt{256
(d-1)+C_{\rm r} (100+4 d (5 C_{\rm r}-38)+(C_{\rm r}-2)^2
d^2)}}\right] \;.{\quad{~}} 
\label{Delta}
\end{eqnarray}
In table \ref{projs} the relevant data are collected for all simple
Lie algebras except ${\rm E}_{8}$ (for which the relevant projectors
have been computed in \cite{Koepsell:1999uj}). In particular, equation
(\ref{Delta}) correctly reproduces the dimensions $\Delta=912$,
$\Delta=351$ for ${\rm G}=$\Exc7, and ${\rm G}=$\Exc6,
respectively. Moreover, the relevant projectors (\ref{E7-projectors}),
(\ref{E6-projectors}) are obtained from (\ref{projectors}).
\begin{table}
\begin{center}
\begin{tabular}{l c c l l r r r}\hline
G & $g^{\vee}$ & $d_\Lambda$ & $\tilde{I}_\Lambda $ & $\Delta$       &
$a_1$   & $a_2$ & $a_3$ \\ 
\hline
${\rm A_r}$ & $r+1$ & $r+1$ &$\frac{1}{2}$    &
$\frac{1}{2}(r-1)(r+1)(r+2)$ &  $\frac{1}{2}$ &  $-\frac{1}{2}$&
$-\frac{1}{2 r}$  \\ 
${\rm B_r}$   & $2r-1$ & $2r+1$& 1    & $\frac{1}{3}r (4 r^2-1)$  &
$\frac{1}{3}$ & $-\frac{2}{3}$ & $0$ \\
${\rm B_r}$   & $2r-1$ & $2^{r}$& $2^{r-3}$    & $2^{r+1}\,r$  &
$\frac{2}{2r-1}$ & $-2^{r-1}\frac{1}{2r-1}$ & $2^{r-1}\,\frac{2r-7}{4
r^2-1}$ \\ 
 ${\rm C_r}$   & $r+1$ & $2r$& $\frac{1}{2}$    & $\frac{8}{3}r
(r^2-1)$  & $\frac{2}{3}$ & $-\frac{2}{3}$ & $-\frac{2}{1+2r}$ \\ 
${\rm D_r}$   & $2r-2$ & $2r$ & 1    & $\frac{2}{3}r (2r^2-3 r +1)$  &
$\frac{1}{3}$ &  $-\frac{2}{3}$ & $0$ \\ 
${\rm D_r}$   & $2r-2$ & $2^{r-1}$ &  $2^{r-4}$   & $2^{r-1}\,(2r-1)$
& $\frac{1}{r-1}$ &  $-2^{r-3}\frac{1}{r-1}$ &
$2^{r-3}\frac{(r-4)}{r\,(r-1)}$ \\ 
 ${\rm G_2}$   & 4 & 7 & 1    & $27$  & $\frac{3}{7}$ & 
$-\frac{6}{7}$ & $-\frac{3}{14}$ \\
 ${\rm F_4}$   & 9 & 26 & 3    & $273$  & $\frac{1}{4}$ &
$-\frac{3}{2}$ & $\frac{1}{4}$ \\
${\rm E}_{6}$ & 12 & 27 & 3 & 351  &  $\frac{1}{5}$  & $-\frac{6}{5}$
& $\frac{3}{10}$ \\ 
${\rm E}_{7}$ & 18 & 56 &6       & 912   & $\frac{1}{7}$ &
$-\frac{12}{7}$ &   $\frac{4}{7}$  \\ 
\hline 
\end{tabular}
\end{center}
\caption{\small
Coefficients for the projector $\mathbb{P}_{{\bf D_1}}$ for the
various algebras. }
\label{projs}
\end{table}

\eject
%

%
\end{document}